\documentclass[aps,pra,twocolumn,superscriptaddress,showpacs,eqsecnum,nofootinbib]{revtex4}
\usepackage{amsbsy,latexsym}
\usepackage{amsfonts}
\usepackage{amssymb}
\usepackage[mathscr]{eucal}
\usepackage{graphics}

\newcommand{\Or}{{\cal O}}
\newcommand{\ord}{\mbox{\tiny$\cal O$}}

\newcommand{\onehalf}{\mbox{\boldmath$\tinyfrac{1}{2}$}}
\newcommand{\rme}{{\rm e}}
\newcommand{\al}{\alpha}
\newcommand{\be}{\beta}

\newcommand{\de}{\delta}
\newcommand{\la}{\lambda}
\newcommand{\id}{\openone}

\newcommand{\tr}{{\rm tr}\,}
\newcommand{\tinyfrac}[2]{\mbox{\footnotesize$ {#1 \over #2}$}}
\newcommand{\smfrac}[2]{\mbox{$ {#1 \over #2}$}}
\newcommand{\ket}[1]{\vert #1 \rangle}
\newcommand{\bra}[1]{\langle #1 \vert}

\newcommand{\D}[3]{\mathfrak D^{(#1)}_{#2 #3}}
\newcommand{\Dc}[3]{\mathfrak D^{(#1) *}_{#2 #3}}

 \newcommand{\rhoN}{\rho^{\otimes N}}
 \newcommand{\eqref}[1]{(\ref{#1})}
 \newcommand{\rbf}{\mathbf{r}}
\newcommand{\Rbf}{\mathbf{R}}
\newcommand{\Vbf}{\mathbf{V}}
\newcommand{\ver}{\vec{r}}
\newcommand{\Young}[6]{
\setlength{\unitlength}{#6pt} \thinlines
\begin{picture}(#3,#1)(0,#5)
\multiput(0,0)(1,0){#4}{\line(0,1){#1}}
\multiput(0,0)(0,1){#2}{\line(1,0){#3}}
\end{picture}
}
\begin{document}
\title{Optimal full estimation of qubit mixed states}

\author{E.~Bagan}
\affiliation{Grup de F{\'\i}sica Te{\`o}rica \& IFAE, Facultat de Ci{\`e}ncies,
Edifici Cn, Universitat Aut{\`o}noma de Barcelona, 08193 Bellaterra
(Barcelona) Spain}

\author{M. A. Ballester}
\affiliation{Department of Mathematics, University of Utrecht, Box
80010, 3508 TA Utrecht, The Netherlands}

\author{R.~D.~Gill}
\affiliation{Department of Mathematics, University of Utrecht, Box
80010, 3508 TA Utrecht, The Netherlands}
 \affiliation{EURANDOM, P.O. Box 513-5600 MB Eindhoven, The
 Netherlands}

\author{A.~Monras}
\affiliation{Grup de F{\'\i}sica Te{\`o}rica \& IFAE, Facultat de Ci{\`e}ncies,
Edifici Cn, Universitat Aut{\`o}noma de Barcelona, 08193 Bellaterra
(Barcelona) Spain}

\author{R.~Mu{\~n}oz-Tapia}
\affiliation{Grup de F{\'\i}sica Te{\`o}rica \& IFAE, Facultat de Ci{\`e}ncies,
Edifici Cn, Universitat Aut{\`o}noma de Barcelona, 08193 Bellaterra
(Barcelona) Spain}
\date{\today}

\begin{abstract}
We obtain the optimal scheme for estimating unknown qubit mixed
states when an
arbitrary number $N$ of identically prepared copies is available. 
We discuss the case of states in the whole Bloch sphere as well as  the
restricted situation where these states are known to lie on the equatorial
plane. For the former case we obtain that the
optimal measurement does not depend on the prior probability
distribution provided  it is isotropic. Although the equatorial-plane case does not have this property
for arbitrary~$N$,
we give a prior-independent scheme which becomes optimal in the asymptotic limit of large~$N$.
We  compute the maximum mean
fidelity in this asymptotic regime for the two cases. We show that
within the pointwise estimation approach these limits can be
obtained in a rather easy and rapid way. This derivation is based
on heuristic arguments that are made rigorous by using  van Trees
inequalities. The interrelation between the estimation of the
purity and the direction of the state is also discussed. In the general case we
show that they correspond to independent estimations whereas for the equatorial-plane
states this is only true asymptotically.
\end{abstract}

\maketitle
\section{Introduction}
Two-state systems or qubits are the 
building blocks of many applications in
Quantum Information. 
Although they are commonly assumed to be in pure states, in real
situations
they are not.
State preparation, processing, quantum channels, etc. are inevitably imperfect, 
which means that any quantum system is, in fact,  in a mixed state. 
The accurate estimation of the parameters that characterize qubit mixed states 
is therefore of utmost
relevance for practical applications. The aim of this work is to find the optimal (most accurate)
scheme to perform this task.

So far, most of the work in state estimation has focused on pure
qubit states~\cite{holevo:book,pure,gillmassar} and fewer quantitative results have
been obtained for qubit mixed states~\cite{cirac, vidal,
ff,keyl,bbm-mixed,zyk-1}. One obvious reason for this is the
greater complexity of the estimation procedure. Whereas pure states are fully
characterized by
just two parameters ---those specifying  a point on the surface of the
Bloch sphere, i.e., a unit vector---
for a mixed state an additional parameter is required to specify its purity, by which
we mean the distance from the center of the Bloch
sphere to the point that represents the state. This brings a theoretical subtlety:
we will need to identify a uniform prior distribution for the purity.
In contrast to the pure-state case where there is a ``natural" uniform probability distribution ---the
invariant measure on the 2-sphere---, for mixed states there
is no unique choice. A uniform distribution must be isotropic (invariant under rotations of the Bloch sphere), but the purity, which is itself invariant, can be distributed according to a
whole  class of functions~\cite{petz, zyk-2}, depending on several
criteria.
Despite this ambiguity, our results turn out to be rather general and, in particular, they
do not depend on the specific choice of an isotropic purity~prior.

In this paper, we assume that we have $N$ identically prepared systems upon
which we can perform generalized measurements.  
{}From their outcomes we can infer the value of the parameters that characterize the state of 
the systems. The quality or accuracy of the estimation is quantified by
the fidelity (to be defined in the next section). 
The  average of the fidelity over the
prior and the outcome
distribution provides a useful summary parameter of the overall quality
of the estimation scheme.
This problem was partially addressed
in~\cite{vidal}. Here we present an alternative formulation that
enables us to apply the approach to new, practically relevant situations
and  find many explicit results.

To be more specific,  we will  study two types of situation:  that of estimating an {\em \`a priori}
completely unknown qubit state and that of
estimating a state that is known to lie on an equatorial plane
of the Bloch sphere. We call the former the 3D case (or just 3D for short), as the state can be represented by any point in the 3-dimensional Bloch sphere. By the same logic, we call the latter~2D. The~2D
case is useful  because in many applications
quantum states can be parametrized by the purity and a phase;
e.g., linearly polarized photons. The~2D case
also exhibits some remarkable theoretical features. For instance,
we will show that while for 3D states the optimal measurement is
essentially unique, independently of the isotropic prior, this is not so for 2D states, though this
feature is recovered in the asymptotic limit of large~$N$.

We will first address the problem from a Bayesian point of view, which will provide 
explicit results for any finite~$N$.
We will also take a steep dive into the asymptotic
regime of  the estimation schemes. It is
clear that unknown states can only be estimated with perfect accuracy in the limit $N\to\infty$. 
The rate at which this perfect determination limit is achieved as $N$ increases is a
very informative parameter. It is useful, e.g., to compare different estimation schemes. If two schemes
have the same rate, we say that they are (asymptotically) equivalent.
The asymptotic behavior is also a central notion in statistics,
where there exists a wealth of results and very powerful techniques~\cite{Matsumoto:holbound,Matsumoto:conditions}. 

Within the
statistical framework, one optimizes over all measurements and
estimators, when the signal state is taken to be fixed. It turns
out, under regularity conditions, that the maximum likelihood
estimator is asymptotically optimal whatever the true signal state.
The mean square error of the estimator gives a measure of the~quality of the
scheme. This error can be related to~the fidelity through the
Fisher information matrix, thus providing a connection with the
Bayesian approach.
In this context, the prior distribution plays a very minor role.
In contrast, within the  Bayesian approach the prior distribution
does play a significant role because, as mentioned above, one is interested in
obtaining an estimation that is optimal on \emph{average}.

Here we present in a fairly comprehensive way the application of
the two approaches to the asymptotic behavior of qubit mixed state
estimation. We will see that both yield the same results. This
fact has important consequences. It tells us that the asymptotic
behaviour of the optimal mean fidelity only depends on the prior
as an average of the optimal {\em pointwise} (i.e., for a fixed state) 
fidelities. 
Second, the Bayesian approach provides an explicit scheme that attains the
pointwise bounds. 
It is worth pointing out that for some restricted
schemes and some priors this might not be the case. For instance,
it is known that a scheme based on fixed local measurements with
the Bures prior distribution~\cite{hubner:excompbures} does not
approach unity at a rate $1/N$~\cite{bbm-mixed}, as a pointwise
approach would indicate. Even more surprising, in this situation
the Bayesian and the Maximum Likelihood estimation give different
asymptotic average fidelities~\cite{bbm-mixed}, in contrast to the
common lore that both estimators should be asymptotically
equivalent,
pointwise.
The non-equivalences here do all have simple explanations.
Pointwise, everything \emph{is} asymptotically equivalent and does
converge at rate $1/N$. However, the convergence is not uniform
or the integrated coefficient of $1/N$ diverges.

This paper is organised as follows. In the next section we introduce
the notation and  main concepts that will be used throughout
this work. In Sec.~\ref{sec-finite} we obtain the optimal
estimation protocol for any number of copies of the state in
both the~3D and the~2D~cases. In Secs.~\ref{sec-asymptotic}  and~\ref{sec-pointwise} we compute
the asymptotic expression of the fidelity from both the Bayesian and the
pointwise approaches, respectively. The derivation of the latter is done through a rather self-contained
presentation since some of the techniques may not be so well known
among physicists. In Sec.~\ref{sec-conclusions} we summarise our
main results. We have relegated many technical details to  the
appendices for the benefit of readers not
interested in technicalities

\section{Preliminaries}\label{sect-preli}
Consider an ensemble of $N$ identically prepared states
$[\rho(\vec r)]^{\otimes N}$, where $\rho(\vec r)$ is a density
matrix with Bloch representation given by
\begin{equation}\label{bloch-3D}
    \rho(\vec r)=\frac{\openone+ \vec{r}\cdot \vec{\sigma}}{2} .
\end{equation}
Here $\vec\sigma=(\sigma^x,\sigma^y,\sigma^z)$, where $\sigma^a$,
$a=x,y,z$, are the usual Pauli matrices and $\vec{r}$ is a point
in the Bloch sphere $\{\vec r : |\vec{r}|\leq 1\}$. We will  drop $\vec
r$ and write simply $\rho$ where no ambiguity arises.

A measurement on $\rhoN$ is represented by a Positive Operator
Valued Measure (POVM). It is defined by a set $O=\{O_\chi\}$ of positive operators such that
\begin{equation}\label{povm}
    \sum_\chi O_\chi=\openone,
\end{equation}
where $\chi $ refers to the various outcomes that can occur. It can
be a discrete or a continuous variable. 

In order to estimate $\rho$ we proceed as follows. We first
perform a measurement on $\rho^{\otimes N}$, from which we obtain
an outcome $\chi $. Based on $\chi $, an estimate for $\rho$ can
be guessed:~$\rho_\chi $.  Its quality is quantified by the fidelity, defined
as~\cite{hubner:excompbures}
\begin{equation}\label{f}
 f(\vec r,\vec R_\chi )=\left(\tr\sqrt{\sqrt{\rho_\chi }\rho\sqrt{\rho_\chi }}\;\right)^2,
\end{equation}
which determines the maximum distinguishability between $\rho$ and
$\rho_\chi $ that can be achieved by any measurement~\cite{fuchs}.
For qubits, Eq.~(\ref{f}) reads
\begin{equation}\label{f-qubits}
 f(\vec{r},\vec{R}_\chi )=\frac{1+\vec{r}\cdot
 \vec{R}_\chi +\sqrt{1-r^2}\sqrt{1-R_\chi ^2}}{2},
\end{equation}
where $\vec{r}$  and $\vec{R}_\chi $ are the Bloch vectors of
the states $\rho$ and $\rho_\chi $ respectively, $r=|\vec r|$ and
$R=|\vec R|$.

In the Bayesian approach the overall performance of the estimation procedure is quantified
by the average fidelity~$F$, hereafter fidelity in short. It is the average of~(\ref{f}) over the prior probability
distribution, which we denote~$d\rho$, and over all possible
outcomes~$\chi $ of a given measurement, namely
\begin{equation}\label{fidelity}
     F=\sum_\chi \int d\rho \, f(\vec{r},\vec{R}_\chi ) p(\chi |\vec{r}),
\end{equation}
where  $p(\chi |\vec{r})$ is the conditional probability of
obtaining outcome~$\chi $ given that the signal state has Bloch
vector~$\vec{r}$.  These probabilities are determined by the
expectation values of the positive operators $O_\chi $, i.e.,~$p(\chi |\vec{r})=\tr [O_\chi  \rho]$. 
Our aim is to maximize~\eqref{fidelity}.

For a given measurement~$O$, there always exists
an  optimal guess or estimator. To prove this,
we first  introduce the four dimensional Euclidean vector
\begin{equation}\label{rmu}
    \mathbf{r}=(r^0,r^x,r^y,r^z)=(r^0,\vec r)=(\sqrt{1-r^2},\vec{r}\,).
\end{equation}
Note that $\rbf\cdot \rbf'=r^0r'{}^0+\vec{r}\cdot\vec{r}\,'$  and
$|\rbf|=\sqrt{\rbf\cdot\rbf}=1$.  With this, the  average fidelity
reads
\begin{equation}\label{fidelity-2}
     F=\sum_\chi \int d\rho \, \frac{1+\rbf\cdot \Rbf_\chi }{2}
     p(\chi |\vec{r}),
\end{equation}
where $\Rbf_\chi =(R^0_\chi ,\vec R_\chi )$ is defined in analogy
to~(\ref{rmu}). A~straightforward use of the Schwarz inequality
gives an upper bound of $F$ that is saturated with the choice
\begin{equation}
  \Rbf_\chi = \frac{\Vbf_\chi }{|\Vbf_\chi |};
  \quad \Vbf_\chi \equiv(V^0_\chi ,\vec V_\chi )\equiv \int d\rho\, \rbf \, p(\chi |\vec{r}),
    \label{vector-1}
\end{equation}
Using \eqref{vector-1}, the maximum fidelity is
\begin{equation}
    F=\frac{1}{2}\left(1+\sum_{\chi }|\Vbf_\chi |\right)\equiv
    \frac{1}{2}\left(1+\Delta \right) .
    \label{fidelity-general}
\end{equation}
Since  the guess \eqref{vector-1} satisfies $|\Rbf_\chi |=1$ and
its first component is non-negative, it \emph{always} gives a
physical state. In fact~\eqref{vector-1} is the best state that
can be inferred and~\eqref{fidelity-general} is the maximum
fidelity that can be obtained given $O$ and the prior~$d\rho$.

In the analysis below, it will prove very convenient to block-diagonalize $\rhoN$ by writing it
in  the basis of
the $\rm SU(2)$ invariant subspaces of $(\onehalf)^{\otimes N}$ [we use boldfaced integers and half-integers to denote the irreducible representations of $\rm SU(2)$],
which are also invariant under the action of the symmetric group~$S_N$ (See App.~\ref{app-decomposition} and also~\cite{vidal,cirac} for details). In
contrast with pure states, for which $\rhoN$ has projection only
in the symmetric $(N+1)$-dimensional subspace of~{\footnotesize$\mathbf{
J}\equiv\mathbf{N\over2}$}, for mixed states $\rhoN$ has also
components in all the lower-dimensional invariant subspaces, 
which, furthermore, occur with multiplicity, $n_j$, greater than one. 
We thus write
 \begin{equation}\label{decomposition'}
    \rhoN=\bigoplus_{j=0,1/2}^{N/2} n_j\rho_{Nj},
\end{equation}
where the lower limit in the direct sum is~$0$ for even~$N$  and~$1/2$
for odd~$N$,
\begin{equation}\label{number-of-repeated}
   n_j=\pmatrix{N \cr N/2-j}\frac{2j+1}{N/2+j+1}
\end{equation}
and
\begin{eqnarray}\label{rhoNj}
     \rho_{Nj}&=&\left(\frac{1-r^2}{4}\right)^{N/2-j}\rho_j,
\end{eqnarray}
with
\begin{eqnarray}
    \rho_j&=&\sum_{m=-j}^{j}
    \left(\frac{1-r}{2}\right)^{j-m}
    \left(\frac{1+r}{2}\right)^{j+m}\times \nonumber \\
    &&U(\vec n)\ket{jm}\bra{jm}U^{\dag}(\vec n).
    \label{rhoj-general}
\end{eqnarray}
Throughout this paper $U(\vec n)$ denotes the $\rm SU(2)$ unitary representation of the
rotation ${\cal R}(\vec n)$  that
takes the unit vector~$\vec{z}$ (pointing along the $z$-axis) into $\vec{n}\equiv \vec r/r$ on the Bloch sphere.
Recall that
\begin{equation}\label{matrix D}
\bra{jm}U(\vec n)\ket{jm'}=\D jm{m'}(\vec n)
\end{equation}
defines the standard Wigner matrices~\cite{edmonds}. Notice that  $\rho_j$ are not
proper
density matrices, since~$\tr \rho_j\not=1$.

For 2D states, the Bloch vector $\vec{r}$ of the state $\rho$ lies
on the equatorial $xy$-plane  of the Bloch sphere, i.e.,
$\vec{r}=r (\cos\theta, \sin\theta,0)$. We are still entitled to use
the decomposition of $\rho^{\otimes N}$ above, but now we write
\begin{eqnarray}\label{rhojj-z-2D}
    \rho_j&=&\sum_{m=-j}^{j}
    \left(\frac{1-r}{2}\right)^{j-m}
    \left(\frac{1+r}{2}\right)^{j+m}\times \nonumber \\
    &&U(\theta) U(\vec x)\ket{jm} \bra{jm} U^\dagger(\vec x)U^{\dagger}(\theta),
\end{eqnarray}
where $\vec x$ is the unit vector pointing along the $x$-axis and
$U(\theta)$ is a unitary representation of a rotation of angle
$\theta$ around the $z$-axis. Note that $U(\vec x)\ket{jm}$ is an
eigenstate of $\vec x\cdot\vec J$ (i.e., of the projection of the total spin operator $\vec J$ along the $x$-axis), since $U(\vec x)$ takes $\vec
z$ into $\vec x$ (i.e., is a rotation of angle $\pi/2$ around the
$y$-axis). Hence, the Bloch vectors of the whole set of states $\{U(\theta)[U(\vec
x)\ket{jm}]\}$ lie on the $xy$-plane, as they should, and $\theta$
is the angle between $\vec r$ and the $x$-axis.

In the basis $\ket{j m}$ the transformation $U(\theta)$ is
diagonal, and substituting~\eqref{rhojj-z-2D} in \eqref{rhoNj}  we
obtain
\begin{equation}\label{rho-theta}
\rho_{Nj}=\sum_{m,m'}\rme^{i (m-m')\theta}\rho^j_{mm'}
    \ket{jm}\bra{jm'},
\end{equation}
where
\begin{eqnarray}
\rho^j_{mm'}&=& \sum_{m''} {\rm d}^{(j)}_{m m''}(\pi/2) {\rm
d}^{(j)}_{m' m''}(\pi/2)
\nonumber\\
&\times&
    \left(\frac{1-r}{2}\right)^{N/2-m''}
    \left(\frac{1+r}{2}\right)^{N/2+m''}
    \label{bar-rho}
\end{eqnarray}
and ${\rm d}^{(j)}_{mm'}$ are the (real) reduced Wigner  matrices~\cite{edmonds}.

\section{Finite number of copies. Bayesian estimator}\label{sec-finite}
In this section we obtain the optimal
POVM  and closed expressions of the fidelity for any number of
copies of the signal state. Although the~3D and~2D cases
look similar, we will show that there are remarkable
differences between them.

\subsection{3D states}\label{subsec-finite-3D}
As mentioned in the introduction, we consider $N$ identical
copies of a quantum state which is chosen according to an isotropic prior distribution
\begin{equation}\label{drho}
   d\rho=w(r)\,dr\,dn ,
\end{equation}
where $dn$ is the invariant measure on the 2-sphere
\begin{equation}\label{dn}
   dn=\frac{d(\cos\theta)\,d\phi}{4\pi}
\end{equation}
and $w(r)$ is normalized such
that $\int_0^1 dr \,w(r)=1$.

Let us start by computing the optimal POVM.
We first notice that because of the block-diagonal form of
$\rho^{\otimes N}$ in~(\ref{decomposition'}) we may  just consider
also block-diagonal POVMs, of the form
\begin{equation}\label{POVM oplus}
  O_\chi  =  \bigoplus_{j=0}^J n_j O_{\chi j}, \quad\mbox{such that}\quad
 \sum_\chi  O_{\chi j}= \openone_j ,
\end{equation}
with no loss of generality. Indeed, for any given POVM $\{O_\chi
\}$, we can always construct a new one, $\{\tilde O_{\chi
j\alpha}\}$, through
\begin{equation}\label{POVM tilde}
\tilde O_{\chi j\alpha}={\openone}_{j\alpha}O_\chi
{\openone}_{j\alpha},
\end{equation}
where ${\openone}_{j\alpha}$ is the identity in the $\bf j$-representation subspace 
and $\alpha$ ($1\le\alpha\le n_j$) labels the different occurances of~$\bf j$  in
the Clebsch-Gordan series of~$(\onehalf)^{\otimes N}$. If  $F$  ($\tilde F$) stands for the maximum
fidelity that can be attained using  $\{O_\chi \}$ ($\{\tilde
O_{\chi j\alpha}\}$), we have $F\le  \tilde F$. This is  readily
seen by noticing that the probability $p(\chi|\ver)=\tr[\rhoN
O_{\chi}]$ is the marginal of $p(\chi j \alpha|\ver)=\tr[\rhoN
\tilde O_{\chi j \alpha}]$, i.e., $p(\chi|\ver)=\sum_{j
\alpha}p(\chi j \alpha|\ver)$, and no marginal can be more
informative than the initial probability distribution. Moreover,
because of~(\ref{decomposition'}), if $\{\tilde O_{\chi
j\alpha}\}$ is  to be optimal, we may obviously  replace $\tilde
O_{\chi j1}$, $\tilde O_{\chi j2}$, \dots, $\tilde O_{\chi jn_j}$
by, say, $\tilde
O_{\chi j1}$, $\tilde O_{\chi j1}$, \dots, $\tilde O_{\chi j1}$ without 
changing the fidelity, which leads us to~(\ref{POVM
oplus}).

It is important to note that~(\ref{POVM tilde}) allows us to view~$j$ 
and~$\alpha$ as the outcome of the measurement~$\{\openone_{j\al}\}$. Therefore, in
Eq.~\eqref{fidelity-general} we will have $n_j |\Vbf_{\chi j}|$
instead of $|\Vbf_{\chi}|$, and an additional  summation over~$j$.
Hence, our goal is to maximize
$|{\mathbf V}_{\chi j}|$ for all pairs $(\chi ,j)$, where
\begin{equation}
\label{Vchij} \Vbf_{\chi j}=\int d\rho\, \rbf \, \tr
(\rho^{\otimes N} O_{\chi j}) .
\end{equation}
The~$j$ outcomes give information about the decomposition of
$\rho^{\otimes N}$ as a direct sum of $\rm SU(2)$ irreducible
components. This, in turn, encodes information about $r$. For instance,
if $r=1$ (pure state), the  probability of obtaining the outcome $j=N/2$ is unity. 
For our purposes, all the information concerning the purity of $\rho$
comes from this source, as we now demonstrate.

Since $V^0_{\chi j}$ is
 invariant under rotations, whereas $ \vec V_{\chi j}$
 transforms as a 3-vector,  we may apply  to
 ${\mathbf V}_{\chi j}$ the rotation
 ${\cal R}^{-1}(\vec n_{\chi j})={\cal R}^{\top}(\vec n_{\chi j})$,
 where $\vec n_{\chi j}=\vec V_{\chi j}/|\vec V_{\chi j}|$, and obtain
 ${\mathbf V}'{}_{\chi j}$, such that its $x$- and $y$-components vanish, i.e.,
$V'{}_{\chi j}^x=V'{}_{\chi j}^y=0$ 
and
 \begin{eqnarray}
V'{}_{\chi j}^z&=&\int d\rho\,\left[ {\cal R}^{\top}(\vec n_{\chi j})
\vec r \right]^z \,\tr(\rho^{\otimes N} O_{\chi j}) \nonumber
\\
&=&\int d\rho\,  r \cos\theta  \,\tr\left(\rho^{\otimes
N}\Omega_{\chi j}\right) \label{V'3},
\\
 V'{}_{\chi j}^0&=&\int d\rho\,\sqrt{1-r^2}\tr(\rho^{\otimes N} \Omega_{\chi j}) ,
 \label{V'0}
 \end{eqnarray}
 where we have defined
 \begin{equation}\label{def Omega}
 \Omega_{\chi j}\equiv U^\dagger(\vec n_{\chi j}) \,O_{\chi j} U(\vec n_{\chi j}),
 \end{equation}
 we have used that $d\rho$ is rotationally invariant, and we have written
 $\vec r=r \vec n$ in spherical coordinates, i.e.,
 $\vec n=(\sin\theta \cos\phi,\sin\theta\sin\phi,\cos\theta)$. Therefore,
$ |{\mathbf V}_{\chi j}|=|{\mathbf V}'{}_{\chi j}| $, and the
maximum fidelity can be computed using ${\mathbf V}'{}_\chi ^j$
instead of~${\mathbf V}{}_\chi ^j$. Hereafter, we drop the
primes and write
\begin{equation}
\Delta_{\rm 3D}=\sum_{\chi j} n_j|{\mathbf V}_{\chi j}|= \sum_{\chi j}
n_j\sqrt{(V_{\chi j}^0)^2+(V_{\chi j}^z)^2}  ,
\end{equation}
where $V_{\chi j}^0$, $V_{\chi j}^z$ are given by~(\ref{V'3})
and~(\ref{V'0}).

Using Eqs.~(\ref{rhoNj}--\ref{matrix D}) and recalling that
$\cos\theta =\D100(\vec n)$, we have
\begin{eqnarray}
V_{\chi j}^z&=&\int_0^1 dr w(r)\,r \sum_{mm'm''} \rho_{jm} \left[\Omega_{\chi j}\right]_{m''m'}\nonumber\\
&\times& \int dn\, \D100(\vec n) \D j{m'}{m}(\vec n)\Dc
j{m''}{m}(\vec n)  ,
\\
V_{\chi j}^0&=&\int_0^1 dr w(r)\,\sqrt{1-r^2}\sum_{mm'm''} \rho_{jm} \left[\Omega_{\chi j}\right]_{m''m'}\nonumber\\
&\times& \int dn\, \D j{m'}{m}(\vec n)\Dc j{m''}{m}(\vec n),
\end{eqnarray}
where the sum over the indexes $m$, $m'$, $m''$ runs from $-j$ to
$j$, and we have defined
\begin{equation}\label{rhojm}
\rho_{jm}=\left({1-r^2\over4}\right)^{J-j}\left({1-r\over2}\right)^{j-m}\left({1+r\over2}\right)^{j+m}.
\end{equation}
The orthogonality relations of the irreducible
representations of $\rm SU(2)$ (Eqs.~(4.6.1) and~(4.6.2) on Page~62 of
Ref.~\cite{edmonds})
enable us to write
\begin{eqnarray}
&& \kern -3em V_{\chi j}^z=\int_0^1 dr {w(r)\,r \over
j(j+1)d_j}\sum_{mm'} mm' \rho_{jm} \left[\Omega_{\chi
j}\right]_{m'm'}, \label{vchi-z-sum}
\\
&& \kern -3em V_{\chi j}^0=\int_0^1 dr {w(r)\,\sqrt{1-r^2}\over
d_j}\sum_{mm'} \rho_{jm} \left[\Omega_{\chi j}\right]_{m'm'},
\label{vchi-0-sum}
\end{eqnarray}
where $d_j=2j+1$ is the dimension of the representation~$\bf j$ of~$\rm SU(2)$.
We readily see that the  $z$- and $0$-components of ${\mathbf
V}_{\chi j}$ are bounded by
\begin{eqnarray}
&& \kern -3em |V_{\chi j}^z|\le{\tr\Omega_{\chi j}\over
d_j}{\displaystyle\max_{m'} |m'|\over j(j+1)}\left|\int_0^1 dr w(r)\,r
\sum_{m} m \rho_{jm} \right|,
\label{inequality max}\\
&& \kern -3em |V_{\chi j}^0|={\tr\Omega_{\chi j}\over d_j}\int_0^1 dr
w(r)\,\sqrt{1-r^2}\sum_{m} \rho_{jm} \label{equality max}.
\end{eqnarray}
Note that all the~$\chi$ dependence has been factored out and~$\Delta_{\rm 3D}$ 
takes the form
\begin{equation}
\label{Delta le} 
\Delta_{\rm 3D} \le \sum_j{n_j}\left(
\frac{\sum_\chi\tr\Omega_{\chi j}}{d_j}
\right)\,\sqrt{(v_{j}^0)^2+(v_j^z)^2}  \ , \label{bound-Delta}
\end{equation}
where $v_j^0$ and $v_j^z$ can be easily worked out
from~(\ref{inequality max}) and~(\ref{equality max}) to be
\begin{eqnarray}
v_j^z&=&\int_0^1 dr{w(r) r\over j+1}\sum_{m=-j}^j m\rho_{jm}\label{coefficient vzj} ,\\
v_j^0&=&\int_0^1 dr\,w(r) \sqrt{1-r^2}\sum_{m=-j}^j\rho_{jm}.
\label{coefficient v0j}
\end{eqnarray}
Eq.~\eqref{def Omega} clearly implies that the factor in
parentheses in~\eqref{Delta le} is unity. Notice that the~$\chi$
dependence has entirely disappeared in the final bound of the
fidelity.

Inequality~(\ref{Delta le}) is saturated iff the only
non-vanishing term of the sum over $m'$ in~\eqref{vchi-z-sum}
corresponds to the maximum value of $|m'|$, namely, $j$. This
implies that $\left[\Omega_{\chi j}\right]_{m'm'}\propto
\delta_{m'j} $ (or the trivial symmetric choice $\delta_{m'-j}$).
An obvious choice that satisfies this condition ---and is independent of $\chi$---  is
\begin{equation}
\label{Omega opt jj} \Omega_{j}= d_j\ket{jj}\bra{jj}.
\end{equation}
The operator $\Omega_{j}$ is a seed of a continuous covariant
POVM, i.e.,
\begin{equation}\label{povm-continuous}
O_{\vec \mu\, j}=U(\vec \mu)\Omega_j U^\dagger(\vec \mu),
\end{equation}
where $\vec\mu$ plays the role of $\chi$. It can be easily
verified that, $\int d\mu\,O_{\vec \mu\,
j}=\id_j$~\cite{holevo:book}, where $d\mu$ (as $dn$) is the
invariant measure over the 2-sphere. This proves that the bound is
attainable. POVMs with a finite number of outcomes can also be 
obtained using the results
in~\cite{bbm-isotropic}.

Having obtained the optimal POVM, Eq.~\eqref{povm-continuous}, it is
straightforward to compute the conditional probabilities
\begin{equation}\label{conditional-probabilities}
   \tr \left(\rho^{\otimes N} O_{\vec \mu j}\right)=d_j\left(\frac{1-r^2}{4}\right)^{J-j}
     \left(\frac{1+ \vec{r}\cdot
    \vec{\mu}}{2}\right)^{2j},
\end{equation}
which will be needed in Sec.~\ref{sec-pointwise}. One
can check that
\begin{equation}\label{povm-identity}
   \sum_j n_j \int d\mu\,  \tr \left(\rho^{\otimes N} O_{\vec \mu
   j}\right)=1,
\end{equation}
as it should be.
The corresponding guesses can be worked out
from~\eqref{Vchij} by simply substituting~$\vec{\mu}$ for~$\chi$. One
can also verify that the angular integration indeed yields the two
terms \eqref{coefficient vzj} and \eqref{coefficient v0j}.

In summary, the fidelity of any optimal POVM can be written as
\begin{equation}\label{implicit vz v0}
\Delta_{\rm 3D}=\sum_j^J n_j   \sqrt{(v^0_j)^2+(v^z_j)^2}.
\end{equation}
This equation along with~\eqref{coefficient vzj} and
 \eqref{coefficient v0j},
provide  a general expression of the maximum fidelity for any
given prior distribution $w(r)$. Unless an explicit expression for
$w(r)$ is given, this is as far as we can get. In
App.~\ref{app-bures} we present closed expressions of the fidelity
for arbitrary~$N$ using the Bures prior.   In the asymptotic limit
$N\to\infty$  however one can derive a compact formula for the
fidelity in terms of the  mean value of~$r$:  $\langle r\rangle=\int_0^1 dr
\,  w(r)\, r$. This will be done in Secs.~\ref{sec-asymptotic}
and~\ref{sec-pointwise}.

Several comments are in order here. Within an optimal scheme, the
purity estimator, 
\begin{equation}\label{R_chij ratio}
R_{\chi j}={|\vec V_{\chi j}|\over |\Vbf_{\chi
j}|}={|v^z_j|\over \sqrt{{v^0_j}^2+{v^z_j}^2}}\equiv R_j, 
\end{equation}
only depends on~$j$ and comes {\em solely} from the measurement represented by the POVM~$\{\id_{j\alpha}\}$~\cite{bmo-purity}. 
All dependence on any other kind of outcome, generically referred to as~$\chi$ [e.g., $\vec\mu$ 
in Eq.~(\ref{povm-continuous})],
has disappeared. This is expected from symmetry grounds: the
parameter~$r$ does not change under~$\rm SU(2)$ transformations and  the optimal purity
guess must thus be a function of $j/N$, as the only $\rm SU(2)$-invariant quantity in this problem is precisely~$j$. Furthermore, since this
measurement ($\{\openone_{j\alpha}\}$) does not alter (on average) the estimation of the orientation~$\vec n=\vec r/r$  of the signal state, the optimal
estimation in the sense of
average fidelity of ({\em a priori}) isotropically distributed mixed states breaks
into two independent estimations: that of the
purity $r$ and that of the  orientation $\vec{n}$ in the Bloch sphere. 
Notice finally that after this measurement, the rest of the protocol, which involves the POVM~(\ref{povm-continuous})  for a fixed~$j$ (or any version of it with a finite number of outcomes), is identical to
the optimal protocol for estimating a pure state $\ket{\vec{n}}$ given $2j$ identical copies of~it~\cite{pure}.

\subsection{2D states}\label{subsec-finite-2D}

In the situation we are about to consider, ${\mathbf V}_{\chi j}$,
defined by~(\ref{Vchij}), still determines the maximum fidelity
through Eq.~(\ref{fidelity-general}), but $d\rho$ is
\begin{equation}
d\rho=w(r)\,dr{d\theta\over2\pi}
\end{equation}
with $\int_0^1 dr\,w(r)=1$. Since $\vec r$ is a 2-dimensional  vector,
we can use a complex notation and write $\vec r\to
r\rme^{i\theta}$. In this notation $\vec V_{\chi j}$ and $\vec
R_{\chi j}$ also become complex numbers. More specifically,
\begin{equation}\label{f0}
    V^0_{\chi j}=\sum_{m}\int_0^1 dr\, w(r)\sqrt{1-r^2} \rho^j_{mm} O^{\chi j}_{mm} ,
\end{equation}
where we have raised the outcome labels $\chi$ and $j$ in $O^{\chi
j}_{mm}$ [or in $\rho^j_{mm'}$, Eq.~(\ref{bar-rho})] to avoid a confusing proliferation
of subindexes; the latter will label  matrix elements, e.g.,
$O^{\chi j}_{mm'}=\bra{jm}O_{\chi j}\ket{jm'}$. Similarly, we have
\begin{eqnarray}\label{f1}
 |\vec V_{\chi j}|&=&\left|\int d\rho\, r \sum_{mm'}  \rme^{i
   (m-m'+1)\theta}
 \rho^j_{mm'} O^{\chi j}_{m'm} \right|\nonumber  \\
  &\le&\int_0^1 dr\, w(r)\, r \sum_{m}
   \rho^{j}_{mm+1} \left| O^{\chi j}_{m+1\,m} \right|,
\end{eqnarray}
where we have used that $\rho^j_{m\,m+1}\ge0$ for all~$r$. The equality in~(\ref{f1}) is attained by choosing the phase of $O^{\chi
j}_{m+1\,m}$ to be independent of~$m$.

The positivity of $O_{\chi j}$ implies that
\begin{equation}\label{positivity}
|O^{\chi j}_{m+1\,m}| \le \sqrt{O^{\chi j}_{mm}}\sqrt{O^{\chi
j}_{m+1\,m+1}}   .
\end{equation}
By choosing $|O^{\chi j}_{m+1\,m}|$ to take its maximum value
in~(\ref{positivity}) we ensure that  $|\vec V_{\chi j}|$ will
also be maximal. So far, the optimization of $V^0_{\chi j}$ and
$|\vec V_{\chi j}|$ can be carried out independently of one
another, since the choices we have to make in order to saturate
the bounds in~(\ref{f1}) and~(\ref{positivity}) do not affect
$V^0_{\chi j}$.  However, we will have to check that they are
compatible with the POVM condition $\sum_\chi O_{\chi
j}=\openone^j$. We will verify this by giving an explicit POVM that
meets all the above conditions.

We now replace $O_{\chi j}$ by its covariant version~$\tilde O_{\chi j\phi}$, defined
in~(\ref{tilde O}) ---in Appendix~\ref{app-covariant-povms-2d} we show that this change does not affect  the average fidelity---
and take the seed (positive) operator $\Omega_{\chi j}$
in~(\ref{covariant POVM 2D}) to be given by $\Omega_{\chi j}=\ket{u_{\chi
j}}\bra{u_{\chi j}}$ (i.e., to be rank one), where
\begin{equation}
\ket{u_{\chi j}}=\sum_m u^{\chi j}_m\ket{j,m}.
\end{equation}
The components $u^{\chi j}_m$ are taken to be real and must satify
\begin{equation}\label{u normalization}
\sum_\chi \left(u^{\chi j}_m\right)^2=\sum_\chi O^{\chi j}_{mm}=1,
\end{equation}
as follows from
\begin{equation}
\tilde O^{\chi j\phi}_{mm'}=\rme^{i(m-m')\phi} u^{\chi j}_m
u^{\chi j}_{m'}  .
\end{equation}
It is important to realize that the vanishing of the off-diagonal
elements in $\sum_\chi\int_0^{2\pi} d\phi/(2\pi)\,\tilde O^{\chi
j\phi}_{mm'}=\id^j$ does not require further conditions
on~$u^{\chi j}_m$. Moreover,
\begin{eqnarray}
\tilde O^{\chi j\phi}_{m+1\,m}&=&\rme^{i\phi} u^{\chi
j}_{m+1}u^{\chi j}_m\nonumber
\\
&=&\rme^{i\phi} \sqrt{\tilde O^{\chi
j\phi}_{mm}} \sqrt{\tilde O^{\chi j\phi}_{m+1\,m+1}},
\end{eqnarray}
hence, this choice saturates both~(\ref{f1})
and~(\ref{positivity}).

Collecting all the pieces and defining  $\Delta_{\rm 2D}=\sum_j n_j
\Delta^{\rm2D}_j$ [recall that $F=(1+\Delta)/2)$], we see that the maximum
fidelity is given by the maximum value of
\begin{eqnarray}
\Delta^{2D}_j&=&\sum_\chi \left\{\left[\sum_m \alpha^{j}_m (u^{\chi j}_m)^2\right]^2\right.\nonumber\\
&+&\left.\left(\sum_m\beta^j_m u^{\chi j}_m u^{\chi
j}_{m+1}\right)^2\right\}^{1/2}, \label{Deltaj}
\end{eqnarray}
where $u^{\chi j}_m$ is constrained by~(\ref{u normalization}) and
$\alpha^j_m$ and $\beta^j_m$ can be read off from~(\ref{f0})
and~(\ref{f1}) respectively:
\begin{eqnarray}
\alpha^{j}_m&=&\int_0^1 dr\,w(r)\sqrt{1-r^2}\rho^j_{mm}\label{al}\\
\beta^{j}_m&=&\int_0^1 dr\,w(r)\,r\,\rho^j_{mm+1}, \label{be}
\end{eqnarray}
With no loss of generality we can take the index~$\chi$ in~(\ref{Deltaj}) to be integer and its maximum value to be less
or equal than the number of distinct values of $\alpha^j_m$
in~(\ref{al}). The symmetry relation ${\rm d}^{(j)}_{mm'}={\rm
d}^{(j)}_{-m'\,-m}$
further implies that $\chi\le{}[d_j/2{}]$, where ${}[\dots{}]$ stands for
integer part. With all the above, maximizing $\Delta_{\rm2D}$, which can
be done for each~$j$ independently, becomes a straightforward
task.

The results of the 3D case may lead us to believe that the optimal
POVM will be independent of the prior $w(r)$. The inspection of the
low $N$ cases gives further support to this belief. For $j\le 5/2$
($N\le 5$) one can  show that the optimal POVM is given by
\begin{equation}\label{u^j_m=1}
u^{j}_m=1  
\end{equation}
for {\em any} prior $w(r)$, where we have dropped the index $\chi$
because it only takes one value here.\footnote{There are also
degenerate solutions of the form $u^{\chi j}_m=\lambda_{\chi j}$
for all $m$, and with $\sum_\chi \lambda_{\chi j}=1$}
However, one can check that for $j\ge3$ the choice~(\ref{u^j_m=1})
is {\em not} optimal for some priors. Take for instance $N=6$ and consider a
prior of the form  $w(r)=(2r/\delta^2) \Theta(\delta-r)$, where
$\Theta(x)$ is the step function [i.e., $\Theta(x)=1$ for $x\ge0$
and~$\Theta(x)=0$ otherwise] and $\delta$ is a positive
number. If $\delta$ is sufficiently small, one can Taylor-expand $\Delta_3$ 
about $\delta=0$ and easily obtain the optimal solution at leading order, 
which does not turn out to be of
the form~\eqref{u^j_m=1}. A straightforward computation
yields~$(\Delta^{\rm opt}_3-\Delta^{\rm
Eq.\protect\eqref{u^j_m=1}}_3)/\Delta^{\rm
opt}_3=A\delta^4+\Or(\delta^5)$, where $A$ is a constant that can
be computed analytically ($A\approx 1.0\times 10^{-3}$).

In spite of this unexpected dependence on
the prior in the 2D~case, there are, however, two features in the
example above that are completely general: (a)~the difference
$\Delta^{\rm opt}_j-\Delta^{\rm Eq.\protect\eqref{u^j_m=1}}_j$ is always 
very small, and (b)~$\Delta^{\rm opt}_j$ is actually different from $\Delta^{\rm
Eq.\protect\eqref{u^j_m=1}}_j$ only for priors that are very
peaked about~$r=0$.
There is a further, very important property:  the POVM defined
by~\eqref{u^j_m=1} is asymptotically optimal (the proof is given in
Appendix~\ref{app:vantrees}).
Hence, for practical purposes, the best one can do is to stick to
the choice~\eqref{u^j_m=1}, for all $j$ and $m$, regardless the
prior knowledge one may have of $\rho$. Though this choice does not
guarantee optimality for small~$N$, it does guarantee that the corresponding fidelity
will differ from the maximum one by a tiny amount (typically less than only one part in a thousand) and, furthermore,
that this difference will decrease to zero as~$N\to\infty$.  

The asymptotically optimal choice~(\ref{u^j_m=1}) amounts to replacing $O_{\chi j}$ by
\begin{equation}\label{tildeOjphi}
\tilde O_{\phi j}=U(\phi)\Omega_j U^\dagger(\phi),
\end{equation}
where $\Omega_j=\ket{u_j}\bra{u_j}$,  $\ket{u_j}=\sum_m
\ket{jm}$, and [hereafter we drop the superindex ``Eq.~(\ref{u^j_m=1})" in $\Delta$, $\Delta_j$, etc.]
\begin{equation}\label{Delta g0j gxj}
\Delta_{\rm2D}=\sum_j n_j \sqrt{\left(v^0_j\right)^2+\left(v^x_j\right)^2} ,
\end{equation}
 where
\begin{equation}
v^0_j=\sum_m \alpha^j_m,\qquad v^x_j=\sum_m \beta^j_m,
\end{equation}
and the analogy with~\eqref{implicit vz v0} is apparent.

We next recall~\eqref{al}, which involves $\tr\rho_{Nj}$. Since the trace is invariant
under rotations, $v^0_j$ can be straightforwardly computed
using~\eqref{rhoNj} and~\eqref{rhoj-general}.
No such simplification exists for~$v^x_j$, as far as we are aware. Proceeding this way we have
\begin{eqnarray}
v^0_j&=&2\sum_{m=-j}^j\int_0^1 dr\, w(r)\left({1-r\over2}\right)^{J-m+{1\over2}}\nonumber\\
&\times&\left({1+r\over2}\right)^{J+m+{1\over2}}    ,\nonumber\\
v^x_j&=& \sum_{m=-j}^j c^j_m\int_0^1 dr\,r\,
w(r)\left({1-r\over2}\right)^{J-m}\nonumber\\
&\times& \left({1+r\over2}\right)^{J+m} \label{vs-2D},
\end{eqnarray}
where the coefficients $c^j_m$ are given by
\begin{equation}\label{cjm}
c^j_m=\sum_{m'=-j}^{j-1} {\rm d}^{(j)}_{m'm}(\pi/2) {\rm
d}^{(j)}_{m'+1\,m}(\pi/2) ,
\end{equation}
as can be read off from~(\ref{bar-rho}).
The sum over $m$ in $v^0_j$ can be easily performed, since it is
just the sum of a geometric series, and yields
\begin{eqnarray}
v^0_j&=&2\int_0^1 dr{w(r)\over r}\Bigg\{\left({1-r\over2}\right)^{J-j+{1\over2}} \nonumber\\
&\times&\left({1+r\over2} \right)^{J+j+{3\over2}}-(r\to-r)\Bigg\}
.
\label{v0j final}
\end{eqnarray}
The sum over $m$ in $v^x_j$, however, is non trivial because of
the coefficients $c^j_m$ and no simple closed formula can be found
but in the asymptotic limit $N\to\infty$.

\section{Asymptotics: Bayesian approach}\label{sec-asymptotic}

\newcommand{\rlprt}{\mathop{\mathrm{Re}} \nolimits}
\newcommand{\imprt}{\mathop{\mathrm{Im}} \nolimits}
\newcommand{\Abs}{\mathop{\mathrm{Abs}} \nolimits}
\newcommand{\Tr}{\mathop{\mathrm{Tr}}\nolimits}
\newcommand{\beq}{\begin{eqnarray} }
\newcommand{\eeq}{\end{eqnarray}}
\newcommand{\av}{\mathbb{E}}
\newcommand{\reals}{\mathbb{R}}
\newcommand{\dfdas}{\stackrel{\text{def}}{=}}
%
\newcommand{\xt}{{\text{\boldmath$X$}}}
\newcommand{\tht}{{\text{\boldmath$\theta$}}}
\newcommand{\htht}{{\text{\boldmath$\hat\theta$}}}

In this section we calculate the asymptotic (large~$N$) expressions of the fidelities
obtained in the previous sections using the Bayesian approach. For~2D they are summarized in~(\ref{Delta g0j gxj}), with the definitions~(\ref{vs-2D}), (\ref{cjm}) and the relation~(\ref{v0j final}). For~3D the maximum fidelity is given by~(\ref{implicit vz v0}), which involves the definitions~(\ref{coefficient vzj}) and~(\ref{coefficient v0j}).
We here present a detailed computation only for~2D. The 3D~case can
be computed in a similar  way and  we just point out the main
differences with~2D.
For simplicity we  consider an even number of copies $N=2n$,
thus $J=n$.

We start by noticing that the coefficients $c^j_m$, defined in
Eq.~\eqref{cjm}, satisfy $c^j_{-m}=-c^j_m$ (which implies
$c^j_0=0$) and, hence,
\begin{eqnarray}
v^x_j&=&\sum_{m=1}^{j} c^j_m\int_0^1 dr\,r\,w(r)
\Bigg\{\left({1-r\over2}\right)^{n-m} \nonumber\\
&\times& \left({1+r\over2}\right)^{n+m} -(r\to -r)\Bigg\} .
\label{gxj before}
\end{eqnarray}
We further note that the dominant contribution to the sum in
$v^x_j$ comes from the region where $m$ is close to its maximum
value~$j$. We can thus replace~$c^j_m$ by the first terms of
its ``Taylor expansion" about $m=j$. It turns out that only the first
two terms, $c^j_m\approx a_j+b_j (m-j)$, contribute at the order
we are interested in. The coefficients~$a_j$ and~$b_j$ are computed
in Appendix~\ref{A and B}. After substituting Eq.~\eqref{cj
approximate} in~\eqref{gxj before} the sum over $m$ gives:
\begin{eqnarray}
v^x_j=
\int_0^1 dr {w(r)\over r}\bigg\{\left(r-{1\over4j}\right)
\left({1-r\over2}\right)^{n-j}\nonumber\\
\times\left({1+r\over2}\right)^{n+j+1} -(r\to -r)\bigg\},
\label{gxj after}
\end{eqnarray}
where we have dropped terms that fall off exponentially as~$n$
goes to infinity.
It is convenient to combine~$v^0_j$ and~$v^x_j$ with the binomial
in~$n_j$ [see Eq.~\eqref{number-of-repeated}]  and define $\bar{v}^0_j$
and~$\bar{v}^x_j$ as
\begin{equation}
\label{}
 \bar{v}^0_j=\pmatrix{2n\cr n-j}v^0_j,\quad \bar{v}^x_j=\pmatrix{2n\cr n-j}v^x_j .
\end{equation}
With this, Eq.~\eqref{Delta g0j gxj} becomes
\begin{equation}\label{Delta gS0j gSxj}
\Delta_{\rm2D}=\sum_j{d_j\over
n+j+1}\sqrt{\left(\bar{v}^0_j\right)^2+\left(\bar{v}^x_j\right)^2}.
\end{equation}
Our goal is to compute the asymptotic behaviour of the above
sum. We do so by first computing  the leading order 
contribution:~$\lim_{n\to\infty}\Delta$. 
We, of course, expect this to be unity, as the optimal guess must certainly lead to a perfect estimation given infinitely many copies. 
The calculation  thus provides a consistency check of the approach and, moreover,  the leading order expression of $\bar{v}^0_j$ and
$\bar{v}^x_j$, which will be later used to compute the next-to-leading
order contribution.

At leading order in $1/n$, we are entitled to use the well
known result
\begin{equation}\label{gaussian}
\pmatrix{2n\cr k}q^k(1-q)^{2n-k}\approx {\exp\left\{
-n{\left({k\over 2n}-q\right)^2\over q(1-q)} \right\}\over
2\sqrt{\pi n q(1-q)}} ,
\end{equation}
which holds for large~$n$. In our case $k=n-j$ and  $q=(1-r)/2$.
Furthermore, we can approximate the gaussian in~(\ref{gaussian})
by the Dirac delta function  $\delta(k-2n q)=\delta(n
r-j)=\delta(r-j/n)/n$. After a straightforward calculation we end up with 
\begin{eqnarray}
\bar{v}^0_j&=&{1\over2n}{w(s)\over s}(1+s)\sqrt{1-s^2}+\ord(1/n),\nonumber\\
\bar{v}^x_j&=&{1\over2n}w(s)(1+s)+\ord(1/n), \label{gS leading}
\end{eqnarray}
where $s=j/n$.

Recalling the derivation of Eq.~\eqref{Delta g0j gxj}, we see 
that 
the optimal guess for the purity only depends on~$j$ and is given by
\begin{equation}\label{R_chij ratio 2D}
R_j={|v^x_j|\over\sqrt{(v^0_j)^2+(v^x_j)^2}}=
{\bar{v}^x_j\over\sqrt{(\bar{v}^0_j)^2+(\bar{v}^x_j)^2}} ,
\end{equation}
in full analogy with~(\ref{R_chij ratio}). 
[The optimal guess for~$\theta$ is given by~$\phi$, Eq.~(\ref{tildeOjphi}).] One readily obtains 
\begin{equation}\label{Rj=j/n}
R_j={j\over n}+\ord(1), 
\end{equation}
as expected. Similarly, it also follows from~\eqref{gS leading} that
\begin{equation}
\sqrt{(\bar{v}^0_j)^2+(\bar{v}^x_j)^2}={1\over2n}(1+s){w(s)\over
s} +\ord(1/n).
\end{equation}
At leading order the sum over~$j$ in~\eqref{Delta gS0j gSxj} can
be replaced by $n \int_0^1 ds$, and $d_j/(n+j+1)\approx
2j/(n+j)=2s/(1+s)$. Hence, at leading order
\begin{equation}
\Delta_{\rm 2D}=  \int_0^1 ds \, w(s)=1,
\end{equation}
and, as it should be, $ \lim_{N\to\infty}F=1$ for {\em any} prior.

We are now ready to  compute the fidelity to next-to-leading order.  
The calculation  can be greatly simplified by noticing that
\begin{equation}\label{D-2d-ineq}
\Delta_{\rm2D}\ge  \sum_{j=0}^n{d_j\over n+j+1}
\left(\bar{v}^0_j\sqrt{1-\xi_j^2}+\bar{v}^x_j \xi_j \right),
\end{equation}
for all $\xi_j$ such that $0<\xi_j<1$ [this is, in reverse, the same argument
that took us from~\eqref{fidelity-2} to~\eqref{fidelity-general}]. 
The bound is saturated iff 
\begin{equation}
(\sqrt{1-\xi_j^2},\xi_j)\propto (\bar{v}^0_j,\bar{v}^x_j)
\end{equation}
for all $j$, namely, iff $ \xi_j=R_j
$. With the leading order choice $\xi_j=j/n$, Eq.~(\ref{D-2d-ineq}) provides a tight bound
at order~$\ord(1/n)$. At next-to-leading order we thus have
\begin{equation}
\label{n.l.o}
\Delta_{\rm2D}
=  \sum_{j=0}^n{d_j\over n+j+1} \left(\bar{v}^0_j\sqrt{1-{j^2\over
n^2}}+\bar{v}^x_j{j\over n}\right),
\end{equation}
where we have  ``linearized" the
square root in~\eqref{Delta gS0j gSxj}, hence overcoming in a very 
simple way the most demanding  part of the calculation. We can now use  the
techniques in Appendix~\ref{app-asymptotic-explicit}  to evaluate the asymptotic value of this sum. 
We obtain
\begin{equation}
    \Delta_{\rm2D}=\left(1-\frac{1}{2n}\right)\int_0^1dr\,w(r)+\ord(1/n) ,
\end{equation}
which implies
\begin{equation} \label{2D-asymp-bayes}
F^{2D}=1-{1\over2N}+\ord(1/N) ,
\end{equation}
independently of the prior~$w(r)$. This result agrees with the
bound derived from the pointwise approach in the next
section.

The very same approach we have outlined  can be applied  to~3D states, we just have to
replace~${v}^x_j$ by~${v}^z_j$  [see Sec.~\ref{subsec-finite-3D} and  Eqs.~\eqref{optimal V0zj},
\eqref{Vzzj=al-be} and \eqref{bej}].  
To next to leading order we have (see Appendix
\ref{app-asymptotic-explicit} for details)
\begin{equation}
    \Delta_{\rm 3D}=\int_0^1
    dr~w(r)\left(1-\frac{3+2r}{4n}\right).
\end{equation}
Recalling that $n=N/2$, the asymptotic fidelity
reads
\begin{equation}\label{F^3D}
    F^{\rm 3D}=1-\frac{3+2 \langle r\rangle}{4N}+\ord(1/N),
\end{equation}
where $\langle r\rangle$ stands for the mean purity over its prior distribution, namely
\begin{equation}
\langle r \rangle\equiv \int_0^1  dr\, w(r)\, r .
\end{equation}

Particularizing~(\ref{F^3D}) to
the Bures distribution,
Eq.~\eqref{drho3}, we have
\begin{equation}\label{bures-3D-asympt}
    F_{\rm Bures}^{\rm 3D}=1-\left(\frac{3}{4}+\frac{4}{3
    \pi}\right)\frac{1}{N}+\ord(1/N) .
\end{equation}


\section{Asymptotics: Pointwise approach}\label{sec-pointwise}

In the Bayesian approach, described in the previous sections, both the measurement strategy
and the estimator (or guess) ---i.e., the estimation scheme--- are so chosen as to
minimize the average fidelity with respect to
a given prior distribution for any~$N$. 
In contrast, in the so called
pointwise approach, to which this section is devoted, one's goal is to 
optimize the performance of a scheme at a \emph{fixed
point}, $\tht_0$, in parameter space (In this section we will denote the parameters
that specify the states by~$\tht$ and the guesses by~$\htht$, as is standard in statistics).

The aim of this section is to present a bound on the quadratic cost, the so called
quantum Cram{\'e}r-Rao bound (QCRB), and its relation to the fidelity. The
QCRB is a matrix inequality which is in general non-attainable.
However there is a related bound that one can expect to be
saturated asymptotically: the Holevo bound. A scheme
that attains this bound is asymptotically optimal
from the pointwise perspective. 

The pointwise approach relies on the fact that for large~$N$ only  
quadratic cost functions become relevant. 
By appropriate algebraic 
manipulations and averaging over the prior distribution
one can compare this approach with the Bayesian one in the asymptotic limit. 
It is proved rigorously in ~\cite{asqinfbd} that the averaged Holevo bound
leads to an asymptotic upper bound to the globally optimal fidelity for
``smooth'' qubit estimation problems, and for ``smooth''
pure state estimation problems. 
(We have a lucky coincidence for qubits, and for pure states, that fidelity 
can be expressed as a quadratic form in the estimation error
of certain parameters of the state.)  One can expect this bound to be 
asymptotically valid in general, but no rigorous proof has been given yet. 

As to whether or not the averaged Holevo bound is
asymptotically saturated: there exist very good heuristic arguments
that this should be true, but no rigorous proof. (Unpublished work
of M.~Hayashi:  for large $N$ the estimation problem can be
approximated, around a point obtained by a preliminary rough estimate, 
by a Gaussian state estimation problem, for which
the Holevo bound is attained by an 
appropriate generalized heterodyne measurement).

In Sec.~\ref{subsec-finite-3D} we derived the optimal global scheme for~3D
states and showed that it is the same for any isotropic prior
distribution. From the previous considerations we expect it also to be
asymptotically optimal in the pointwise sense. We will show that this is 
indeed the case,
since the optimal fidelity does coincide asymptotically with the averaged 
Holevo bound.

For~2D states the situation is more complex. Recall that the scheme 
defined by~(\ref{u^j_m=1}) is not
optimal for arbitrary~$N$ and general isotropic priors.
Nevertheless, Eq.~\eqref{2D-asymp-bayes} also coincides with
the averaged Holevo bound. This comes close to a proof of the asymptotic optimality
of the scheme. A rigorous proof (see Appendix~\ref{app:vantrees})
can be derived from the van Trees
inequality~\cite{vantrees} (the same inequality is used to get
the more general results in  \cite{asqinfbd}). Thus our approximate solution
(\ref{u^j_m=1}) is asymptotically optimal both from the global and from the
pointwise points of view.

Both the 3D and the 2D cases confirm the conjectures that the averaged
Holevo bound is a sharp asymptotic bound for fidelity, and that the
global optimal scheme is also asymptotically optimal in the pointwise sense.
Global asymptotic optimality does not depend on the prior or on non-local
features of the figure-of-merit.

Before stating the main results, we need to introduce a bit of notation.
Let~$\rho$ be a density matrix parametrized by~$\tht\equiv(\theta_1,\theta_2,\dots,\theta_p) \in \Theta
\subset \reals^p$, where $p$ is the number of
parameters.\footnote{In the~3D case~$p=3$, $\tht=(r,\theta,\phi)$
and~$\Theta={}[0,1]\times{}[0,\pi]\times{}[0,2\pi)$. In the~2D case~$p=2$,
$\tht=(r,\theta)$ and~$\Theta={}[0,1]\times{}[0,2\pi)$.}  Just as in the previous sections,
let us assume we perform a generalized measurement~$O$ on an arbitrary 
state~$\rho(\tht)$. Recall that such measurement is represented by a POVM~$O=\{O_\chi\}$, 
where~$\chi\in\Omega$
labels the various outcomes. Let~$\htht_\chi$ be the
estimate (or guess) of~$\tht$ based on the outcome~$\chi$, i.e.,  $\htht$ is
a mapping from the outcome set~$\Omega$ to the parameter space~$\Theta$:
\begin{eqnarray}
    \nonumber
    \htht:\Omega&\rightarrow & \Theta\\
                  \chi&\mapsto&\htht_\chi.
\end{eqnarray}
A natural way of quantifying the performance of an estimator $\htht$ and a
measurement $O$ at a point $\tht_0$ is provided by the mean square error
matrix (MSE) defined by the matrix elements 
\beq
   \kern-2.2em V_{\al\be}(\tht_0,\htht) &\equiv& \av_{\tht_0}[(\hat\theta_\alpha-\theta_{0\alpha})
    (\hat{\theta}_\beta-\theta_{0\beta})]\nonumber \\
    &=&\sum_{\chi\in \Omega} p(\chi|\tht_0) ~(\hat{\theta}_{\chi\alpha}-
    \theta_{0\alpha})(\hat{\theta}_{\chi\beta}-\theta_{0\beta}),
\eeq 
where the dependence on~$O$ is understood to simplify the notation and, naturally, $p(\chi|\tht_0)=\tr[\rho(\tht_0)O_\chi]$. In the remaining sections of the paper~$\av_{\tht_0}[f]$ stands 
for the expectation value of~$f$
with respect to the probability distribution $p(\chi|\tht_0)$.

An estimator is said
to be \emph{locally unbiased}  (LU) at~$\tht_0$~if
\begin{equation}
    \av_{\tht_0}[\htht]=\tht_0, \qquad \partial_\alpha\av_{\tht}[\hat{\theta}_\beta]\Big|_{\tht=\tht_0}
    \kern-.75em=
    \delta_{\alpha \beta},
\end{equation}
where $\partial_\alpha$ is shorthand for~$\partial/\partial\theta_\alpha$.
Intuitively, these conditions mean that, on average, the estimator
is close to the truth in a small neighborhood of~$\tht_0$. When
these conditions are satisfied for all possible values of~$\tht_0$, 
the estimator is said to be \emph{uniformly unbiased},
or, simply, \emph{unbiased}.
LU estimators play a fundamental role in the pointwise approach.

The Fisher information matrix (FI) is defined as 
\beq I_{\alpha
\beta}(\tht)&\equiv&\av_{\tht}\left[
\partial_\alpha\ln p(\chi|\tht)\;
\partial_\beta\ln p(\chi|\tht) \right]\nonumber \\[.5em]
&=&\sum_{\chi \in \Omega}\frac{\partial_\alpha
p(\chi|\tht)\;\partial_\beta p(\chi|\tht)}{p(\chi|\tht)}. 
\eeq
Note that the FI depends on a specific measurement~$O$, through the probabilities~$p(\chi|\tht)$.

With the above few definitions we can already give a first important result: 
The  Cram\'{e}r-Rao bound (CRB). It states that the MSE of an estimator $\htht\,$~LU
at~$\tht_0$ is lower bounded by the inverse of  the  FI, 
namely,
\begin{equation}
V(\tht_0,\htht)\geq I(\tht_0)^{-1}.
\end{equation}
In spite of  its fundamental character, the CRB has the drawback that the bound it provides refers to
a particular measurement, not necessarily optimal. To go around this difficulty, some new definitions are required

The symmetric logarithmic derivative (SLD), denoted by~$\la_\al(\tht)$ (recall that $\al=1,2,\dots,p$),
is defined as the self-adjoint matrix
that satisfies
\begin{equation}
\partial_\alpha \rho(\tht) = \frac{\rho(\tht) \lambda_\alpha(\tht)+
\lambda_\alpha(\tht)\rho(\tht)}{2}.
\end{equation}
The SLDs for the~2D and 3D~cases (2D~and 3D~models in pointwise terminology) are given in Appendix~\ref{sec:applambdas}.
With this we can now define the quantum Fisher information matrix
(QFI) as
 \beq
 H_{\alpha \beta}(\tht)=
 \rlprt \tr \rho(\tht)\lambda_\alpha(\tht) \lambda_\beta(\tht).
 \eeq
E.g., for the two models studied in this paper the QFIs are
 \beq
 \kern-1em
 H_{\rm 3D}\!\!=\!\!\left( \begin{array}{@{}ccc@{}}
\displaystyle\frac{1}{1-r^2}&0&0\\
0&r^2&0\\
0&0&r^2 \sin^2\theta
\end{array}\right)\!\!;
 H_{2D}\!\!=\!\!\left( \begin{array}{@{}cc@{}}
\displaystyle\frac{1}{1-r^2}&0\\
0&r^2
\end{array}\right)\!\! .
 \eeq

The second important result of this section, due to Braunstein and Caves~\cite{BraunsteinCaves:PRL}, states that 
for a given model all FIs are bounded from above
by the QFI, i.e.,
\begin{equation}
    \label{eq:IleqH}
    I(\tht_0)\leq H(\tht_0)\quad\mbox{for all $\{O_\chi\}$},
\end{equation}
{}from which it immediately follows the QCRB:
\beq
    V(\tht_0,\htht)\geq H(\tht_0)^{-1}  \quad\mbox{for all $\{O_\chi\}$}.
\eeq
Although these bounds are measurement-independent \hbox{---they} depend only on the signal states and the geometric properties of
the space they belong to--- they have the drawback of not being always attainable. 

We have seen above that $H(\tht_0)$  provides information on how small the variance of an
estimator can be at~$\tht_0$. There is still another remarkable property of the QFI that we will need below: its direct relation to the fidelity~\cite{hubner:excompbures}.
Indeed, from its definition [see
Eq.~\eqref{f}],
\begin{equation}
    f(\tht_1,\tht_2)=\left(\tr\sqrt{\sqrt{\rho(\tht_1)}\rho(\tht_2)\sqrt{\rho(\tht_1)}}\;\right)^2 ,
\end{equation}
one obtains
\begin{equation}\label{H from f}
    f(\tht_0,\tht_0+\delta\tht)=1-\frac{1}{4}
    H_{\alpha\beta}(\tht_0)\delta\theta_\alpha \delta\theta_\beta
 +\dots ,
\end{equation}
where the components of $\de\tht$ are assumed to be small (neighboring states). 
Given
a scheme, characterized by $(\{O_\chi\},\htht)$, the average of the fidelity over all possible outcomes is
\begin{eqnarray}
 \kern-2em   \av_{\tht_0}f(\tht,\htht)&=&\sum_{\chi\in\Omega} \tr[\rho(\tht_0)O_\chi]~f(\tht_0,\htht_\chi)
    \nonumber
    \\ \label{eq:fav} &=& 1 - \frac{1}{4}\Tr H(\tht_0)V(\tht_0,\htht)+\dots\, .
    \end{eqnarray}
Our aim is, therefore, to minimize the cost
\begin{equation}
    \label{eq:cost}
    \tr H(\tht_0)V(\tht_0,\htht).
\end{equation}
An optimal measurement, $O_{\rm opt}$, is thus the one that minimizes~(\ref{eq:cost}).

The formalism and results presented so far are completely general and apply to
any model, i.e., to any family of states $\rho(\tht)$.
We now need to introduce the so called $N$-copy model. It is defined by the set of density matrices
$ \rho^N(\tht)$
of the form
\begin{equation}
    \rho^N(\tht)=[\rho(\tht)]^{\otimes N} .
\end{equation}
The ``original" family, $\rho(\tht)$, is sometimes referred to as the single-copy quantum model. Naturally,
we can talk about the variance or MSE of an estimation of the
$N$-copy model, which we denote by $V^N(\tht_0,\htht)$. It is not hard to convince
oneself that the cost~Eq.~\eqref{eq:cost} of the optimal
scheme necessarily scales as~$1/N$, for large 
enough~$N$.\footnote{Just consider a scheme consisting of~$N$ identical
measurements on each copy~$\rho(\tht)$. By definition the cost of the optimal scheme
is less than or equal to the cost of the former, which obviously scales as~$1/N$. This sets a bound
on the cost of the latter
that also scales as~$1/N$.} It is well-known in
classical statistics~\cite{bickeldoksum:book} that under some
regularity conditions the maximum likelihood (ML) estimator is
asymptotically unbiased at $\tht_0$ and its MSE is equal to
$I^{N}(\tht_0)^{-1}$, i.e., the ML estimator achieves the
CRB asymptotically. It follows that  for an optimal measurement 
$\tr H(\tht_0) I^N(\tht_0)^{-1}$ provides an attainable bound to the cost and it will 
scale as~$1/N$ asymptotically.
This lower bound on~(\ref{eq:cost}) can be expressed as
 \begin{equation}\label{trace without =}
  {\tr H(\tht_0)  \bar I^N(\tht_0)^{-1}\over N }+\ord(1/N),
\end{equation}
where $\bar I^N=I^N/N$ is called the normalized FI.
Likewise, for the asymptotic fidelity we have
 \beq \label{eq:asympfav}
 \av_{\tht_0}f^N(\tht_0,\htht_{\rm ML})&=& 1
  - \frac{\tr H(\tht_0) \bar I^N(\tht_0)^{-1}}{4 N }\nonumber \\
&&+\ord(1/N).
 \eeq
which means that our optimization problem amounts to
finding a mea\-sure\-ment $O_{\rm opt}$ that
minimizes $\tr H(\tht_0)\bar I^N(\tht_0)^{-1}$. We next present a powerful mea\-sure\-ment-independent bound to this expression; the so called Holevo bound.

Let $G$ be a positive semi-definite matrix
and
\begin{equation}\label{def of C^N}
C^N_{\tht_0}(G) =\min_{\mbox{\tiny $\displaystyle
\begin{array}{c}
\{(O\ {\rm on}\ \rho^N, \htht)\}\\
\mbox{LU at $\tht_0$}
\end{array}
 $}}
\tr G
V^{N}(\tht_0,\htht),
\end{equation}
where the minimization is over all pairs $(O,\htht)$ of measurements on~$\rho^N(\tht)$ and estimators for which the latter is LU at $\tht_0$ (the unbiasedness of an estimator depends on the measurement through its outcome probability distribution).
Eq.~(\ref{def of C^N}) is relevant to the problem we are dealing with because its right hand side
can be shown to give the $1/N$ term in~(\ref{trace without =}) and~(\ref{eq:asympfav}) if $G=H(\tht_0)$.
In Ref.~{\cite{holevo:book}} Holevo  proved the following bound: 
\beq
    \label{eq:holevobound}
    C^1_{\tht_0}(G) &\geq&C^H_{\tht_0}(G),
\eeq 
where 
\beq\label{C^H}
    C^H_{\tht_0}(G)&=&\min_{\xt\in \Xi_{\tht_0}}
  \bigg\{  \tr G \rlprt Z[\xt]\nonumber\\
    &+& \tr \left| \sqrt{G} \imprt Z[\xt]\sqrt{G}\,\right|  \bigg\}.
\eeq
In this expression $\xt=(X_1,X_2,\dots, X_p)$ are hermitian  matrices satisfying the following relations
\begin{eqnarray}
\tr \rho(\tht_0)X_\alpha
    &=&0\label{eq:Xcond1},\\
\tr \partial_\alpha \rho(\tht_0)X_\beta &=&\delta_{\alpha \beta}.
\label{eq:Xcond2}
\end{eqnarray}
The minimization in~(\ref{C^H}) is over the set $\Xi_{\tht_0}$ of all such~$\xt$.
Finally,  $Z[\xt]$  is the $p\times p$ matrix whose elements are given by
\begin{equation}
Z_{\al\be}[\xt]= \tr \rho(\tht_0) X_\alpha X_\beta  .
\end{equation}

Although the Holevo bound~(\ref{eq:holevobound}) is not attainable but for a few simple exceptions, 
unpublished work by M. Hayashi shows that
it is {\em asymptotically}  attainable,  i.e.,
\beq \label{eq:asympattholevobound}
\lim_{N\to\infty}N C^N_{\tht_0}(G) =C^H_{\tht_0}(G),
 \eeq
as previously mentioned in this section. It is important to point out here that practical use of Hayashi's construction would require a two-step measurement in order to saturate the bound. This is necessary because the optimal measurement and LU estimator at $\tht_0$ depend themselves on $\tht_0$, which we do not know beforehand. To overcome this difficulty, one takes an asymptotically vanishing fraction
of copies, say $\sqrt{N}$, and makes an initial estimate of the
parameter $\htht_{\rm ini}$. Then, on the remaining copies one
performs the measurement that is optimal at $\htht_{\rm ini}$. 
Therefore, (\ref{eq:asympfav}) and~(\ref{eq:asympattholevobound}) lead us to expect that the optimal  asymptotic fidelity is given by
\begin{equation}
    \av_{\tht_0}f^N(\tht_0,\htht_{\rm ML})=1-\frac{1}{4N}C^H_{\tht_0}[H(\tht_0)]+\ord(1/N).
\end{equation}
We next apply these results to the~3D and~2D ~models.

 \subsubsection{Holevo bound for the 3D case}
 
 In this case $p=3$ and it is not hard to show (see Appendix~\ref{sec:applambdas})
 that there is only one ``vector" of matrices $\xt=(X_r,X_\theta,X_\phi)$ in $\Xi_\tht$ and no minimization
 is thus required in~(\ref{C^H}).
 The Holevo bound is straightforwardly computed to be
 \begin{equation}\label{C^H 3D}
 C^H[H(\tht_0)]=3+2r,
\end{equation}
and~(\ref{eq:asympfav}) ~becomes
\beq\label{av^3D}
\av_{\tht_0}f^N(\tht_0,\htht_{\rm ML})= 1 - \frac{3+2 r}{4
N}+\ord(1/N).
 \eeq
Furthermore, we expect this result  to hold
regardless on whether  the ML estimator or the optimal guess is used. 
This implies that for a ``well behaved" prior, one should have~(\ref{F^3D}) by simply averaging~(\ref{av^3D}),
and we
re-obtain the result of the the preceding section,
which was computed using the Bayesian approach, with much less effort.
Eq.\ (\ref{av^3D}) was  also obtained by Matsumoto and Hayashi \cite{Matsumoto:holbound} with an estimation strategy similar to the one  developed in Section \ref{subsec-finite-3D}.

\subsubsection{Holevo bound for the 2D case}

In the 2D model the SLDs satisfy
 \beq\label{eq:qccond}
 \imprt \tr \rho(\tht_0)\, \lambda_\alpha(\tht_0)\, \lambda_\beta(\tht_0)=0.
\eeq 
It is not difficult to check that in this situation the
QCRB is asymptotically attainable,\footnote{A
theorem by Matsumoto~\cite{Matsumoto:holbound} states that the QCRB
is asymptotically attainable if and only if~(\ref{eq:qccond})
holds.}
i.e.,
\beq
C^H_{\tht_0}(G)&=&\tr GH(\tht_0)^{-1} .
\eeq 
Indeed, the choice
$X_\alpha=\sum_{\beta}H^{-1}_{\alpha \beta}(\tht_0)\lambda_\beta(\tht_0)$ achieves
this. Hence $C^H_{\tht_0}[H(\tht_0)]=2$ and
\begin{equation}
\av_{\tht_0}f^N(\tht,\htht_{\rm ML})= 1 - \frac{1}{2 N}+\ord(1/N),
\end{equation}
{}from which~(\ref{2D-asymp-bayes}) follows for ``well behaved" priors.  
This strongly supports the claim that the~2D measurement scheme
defined by~Eq.~\eqref{u^j_m=1} is indeed asymptotically optimal. The Appendix~\ref{app:vantrees}
contains the rigorous proof.

\section{Conclusions}\label{sec-conclusions}

We have presented a detailed analysis of the optimal estimation of
qubit mixed states given a number $N$ of identical copies.
Our results 
apply to arbitrary $N$, finite or asymptotically large.

For general states (3D) we have obtained that the structure of the
optimal measurement  is based on the decomposition of the signal
states in irreducible blocks under the action of the symmetric group. 
The scheme is essentially unique,
valid for \emph{any} isotropic prior distribution and \emph{any}
number of copies. This optimal scheme has the nice property that
it can be regarded as two independent  protocols performed sequentially: that for estimating
the purity~$r$ of the state and that for estimating its orientation~$\vec{n}$ in
the Bloch sphere. It turns out that the estimation of the purity
only exploits rotationally invariant properties of the signal states, and
a measurement of the Casimir operator~$\vec J\,{}^2=j(j+1)\sum_\alpha\openone_{j\alpha}$ is optimal.
In other words, the estimate of~$r$ only depends on~$j$, which
characterizes the~$\rm SU(2)$ invariant subspaces. 
This should not come as a surprise since the  purity itself is rotationally invariant
and so are the priors considered here. 
The estimation of the orientation is formally equivalent to
pure state estimation with $2j$ copies. As an illustration of
our procedure, we have obtained closed expressions of the fidelity
for the particularly important Bures prior. Results for other priors can be easily
obtained with the techniques presented here.

In 2D, if one wants to do precisely optimal estimation for any~$N$, there is a subtle interplay between the estimation of the purity and the estimation of
the phase and  they are no longer independent, although they are {\em asymptotically}~so.  Also contrasting with 3D is
that the structure of the optimal POVM depends on the prior. The roots of this
unconventional behavior  lie in the different group
structure of~2D states.  Here the relevant group is~$\rm U(1)$ [instead of~$SU(2)$]
and~$j$ is not the only invariant; the magnetic number~$m$ is also invariant under~$\rm U(1)$. 
Actually, the interplay purity-phase can be traced back to this symmetry property.  
In spite of these difficulties, we have
reduced the problem of obtaining 
the optimal
POVM for any isotropic prior to a
rather trivial maximization problem [recall Eq.~(\ref{Deltaj})].
We have also obtained  a prior
independent POVM that is indistinguishable from the optimal one 
for any practical purposes. Furthermore, it separates purity and phase estimation exactly for all~$N$ and 
is asymptotically optimal.

The asymptotic behaviour of the estimation procedure has also been
a central issue of our work. The asymptotic fidelity in 3D has
the simple form  $F=1- (3+2\langle r\rangle)/(4N)$, where $\langle
r\rangle$ is the mean purity with respect to the prior.
This result is proved here for isotropic priors within our Bayesian
approach. 
It is worth emphasizing that so far the
asymptotic expression was only known for the particular case of
the Bures prior~\cite{bbm-mixed}.  In 2D, the asymptotic fidelity
computed with the fixed POVM described above is simply $F=1- 1/(2N)$,
independently of the prior.

We have studied the asymptotic behavior also from the pointwise
approach, which is
far more common among statisticians. The main advantage of the pointwise
approach over the Bayesian one is that it provides  
bounds on  the asymptotic mean square error
(as well as on any other quadratic loss function)
that can be easily computed.
These
bounds correspond, by second order expansion of the figure-of-merit, 
to bounds on the average fidelity which can be shown 
to be rigorous in many cases (\cite{asqinfbd}),
including those studied in this paper.
The drawback of the approach is that though one can heuristically expect these
bounds to be asymptotically sharp, and one can propose two-stage measurement
schemes which can be hoped to do the job, a lot of hard work is needed 
in each case to prove that they can be achieved.
In contrast with the
3D case where all the  results we have worked out from the Bayesian approach are rigorous, the optimality in the asymptotic regime of the 2D estimation scheme defined by~(\ref{u^j_m=1}) or~(\ref{tildeOjphi})  required some
further work. 
Here we used the pointwise approach to fill the gap.
The application of the van Trees
inequality~\cite{vantrees} to 2D in Appendix \ref{app:vantrees}
yields the asymptotic bound on the fidelity in a
particularly elegant and straightforward way. 
In turn, this bound provides the optimality proof.

Altogether, the fact that the results obtained from the~pointwise approach coincide with
those derived from~the Bayesian framework give further strong support
for the heuristic principle that the averaged lower bound from the
pointwise approach is an asymptotically sharp lower bound for the
global approach; and moreover that the chosen prior distribution
and to a lesser extent, figure-of-merit, has asymptotically little
impact on the behaviour of the solution.

There are two extensions of our work that can be readily
addressed. Here, we have considered the full estimation of a qubit
mixed state, however for some applications only partial knowledge
of the state, such as its purity or its orientation, may be
required. The techniques developed in this work can be easily
adapted to these situations (see \cite{bmo-purity} and
\cite{us-japon}). A second line of work concerns the use of more
realistic measurements, in particular those that can be
implemented with current technology. In this work we have
considered the most general measurements allowed by Quantum
Mechanics. They yield the maximum theoretical accuracy that can
possibly be achieved, and thus provide a bound (and a measuring rod) for
the  accuracy of any other estimation scheme. However, they
involve joint operations on the whole sample of states  that in
general are difficult to implement in a laboratory. It is thus of
great practical relevance to study schemes based on local von
Neumann measurements. Preliminary
results, were presented in~\cite{bbm-mixed}. There, it was found
that, for some tomographic schemes, the rate at which the fidelity
approaches unity for a Bures prior distribution is $1-F\sim
1/N^{3/4}$,  i.e., there is a \emph{qualitative} difference with
the optimal measurements. Present work in progress suggests that
by using classical communication the precision rate can be similar to
the optimal collective scheme  $1-F\sim 1/N$, but the coefficient
of the $1/N$ term is strictly larger than the optimal one, 
and corresponds to the result from the pointwise approach obtained
in~\cite{gillmassar}.

\begin{acknowledgments}

We thank M.~Baig for his  collaboration at early stages of this
work.  We acknowledge financial support from Spa\-nish Ministry of
Science and Technology project BFM2002-02588, CIRIT project
SGR-00185, Netherlands Organization for Scientific Research NWO project 613.003.047,
the European Community projects QUPRODIS  contract no.
IST-2001-38877 and RESQ contract no IST-2001-37559.
E.~B. thanks EURANDOM for hospitality.
\end{acknowledgments}


\appendix


\section{Block-diagonal form of~{\boldmath$\rhoN$}}\label{app-decomposition}
One may use the symmetric group~$S_N$ to write~$\rhoN$ in the 
block-diagonal form~(\ref{decomposition'}), 
much in the same way as it is used to obtain the~$\rm SU(2)$ Clebsch-Gordan decomposition
\begin{equation}\label{appA:C-G series}
\left(\mbox{\boldmath${1\over2}$}\right)^{\otimes N}=\bigoplus_{
j=0,1/2}^{J} n_j \mathbf j \quad (J=N/2)
\end{equation}
(the multiplicity, $n_j$, is computed in
Appendix~\ref{app multiplicity}). 
However, at variance with the $\rm SU(2)$ case, where all Young frames have a single row, we here
must also consider those with two rows, because
\begin{equation}\label{appA:det rho}
\det \rho={1-r^2\over4}  
\end{equation}
(instead of unity). Hence, each two-box column of a frame contributes a multiplicative factor~$\det\rho$.

With this observation, one can easily obtain the expression of the blocks~$\rho_{Nj}$ as follows.
A generic Young frame with $N$ boxes has the shape
\begin{eqnarray}
& 
\overbrace{ \Young2334{1.25}9\raisebox{-.3em}{ \dots}
\Young2312{1.25}9}^{{N\over2}-j\ {\rm columns}}
\kern-.16em\overbrace{\Young1234{.25}9\cdots\Young1212{.25}9}^{2j\
{\rm columns}}  .
& \label{appA:young tableaux}\\
&&\nonumber
\end{eqnarray}
Each of the $N/2-j$ double columns gives a factor $\det\rho$. The
remaining $2j$ single columns correspond to a  fully symmetric tensor on which $\rm SU(2)$ acts irreducibly. In the basis of the irreducible subspace of the representation~$\bf j$, this tensor can be written as the matrix which we
denote by~$\rho_j$. Hence
\begin{equation}\label{appA:rhoNj}
     \rho_{Nj}=\left(\frac{1-r^2}{4}\right)^{J-j}\rho_j.
\end{equation}

We now note that for $\vec r=r\vec z$
the matrices $\rho^{\otimes N}$, $\rho_{Nj}$ and  $\rho_j$, are
all them diagonal and can thus be obtained without much effort. The result
is
\begin{equation}\label{appA:rhoj-z}
    \rho_j
    =\sum_{m=-j}^{j} \left(\frac{1-r}{2}\right)^{j-m}
    \left(\frac{1+r}{2}\right)^{j+m}\ket{jm} \bra{jm}.
\end{equation}
For arbitrary~$\vec r$ covariance implies
\begin{eqnarray}\label{appA:rhoj}
    \rho_j&=&\sum_{m=-j}^{j}
    \left(\frac{1-r}{2}\right)^{j-m}
    \left(\frac{1+r}{2}\right)^{j+m}\times \nonumber \\
    &&U(\vec n)\ket{jm}\bra{jm}U^{\dag}(\vec n).
\end{eqnarray}
Notice that, in spite of what the notation might suggest,
the matrices $\rho_j$ are not proper density matrices, as \mbox{$\tr
\rho_j\not=1$.}


\section{The multiplicity of the representation~$\bf j$}\label{app multiplicity}
Using Young tableaux techniques, there is a simple way to compute the multiplicity
$n_j$, (\ref{number-of-repeated}), with which the representation~${\mathbf j}$  shows up in the
Clebsch-Gordan decomposition of $(\onehalf)^{\otimes
N}$ (this tensor product is denoted by $\protect\Young1212{0}6{}^{\otimes N}$ in the present context).

  The Young frame in~(\ref{appA:young tableaux}) can  be denoted by
$\lambda={}[\lambda_1,\lambda_2]={}[N/2+j,N/2-j]$ (this is a
standard notation where~$\lambda_k$ is the number of boxes in the $k$-th row
of the frame). This very same frame~(\ref{appA:young tableaux}) is equivalent to a single row of~$2j$ boxes, i.e., to $[2j]$, which denotes
the representation~$\bf j$ of $\rm SU(2)$.

The recipe for computing $\rm SU(2)$ Clebsch-Gordan decompositions~\cite{pdg} applied to $\protect\Young1212{0}6{}^{\otimes N}$ amounts to the following. First label~$N$ boxes each with an integer number from~$1$ to~$N$. Then, starting with box number one and proceeding sequentially, build (and keep account of) all possible Young tableaux  
such that (i)~they have at most two rows and (ii)~the full sequence of integers formed by
reading right to left in the  first row and then in the second is 
{\em admissible}.\footnote{A sequence of integers $p,q,r\dots$ is admissible if at any point in
the sequence at least as many $1$'s have occurred as $2$'s, at least as many $2$'s have occurred
as $3$'s, etc. }
The number of occurrences of~(\ref{appA:young tableaux}) is precisely~$n_j$.   But the very same recipe gives us all {\em standard}
Young tableaux\footnote{A Young tableaux is said to be {\em standard} if  its labels increase from
left to right along the files and from top to bottom along the
columns.} of shape $\lambda={}[N/2+j,N/2-j]$. Hence $n_j$ equals the number, $f_\lambda$, of such tableaux.

Recalling the Frobenius
determinantal formula~\cite{ham},
\begin{equation}
f_\lambda= N! \left|\kern-.2em\left|{1\over
\lambda_k-k+l}\right|\kern-.2em\right|,
\end{equation}
we get
\begin{equation}
n_j=N!\left|\protect\matrix{{1\over N/2+j}&{1\over N/2+j+1} \cr
{1\over N/2-j-1}  &  {1\over N/2-j} }
 \right| .
\end{equation}
This determinant is readily seen to
give~(\ref{number-of-repeated}).


\section{Closed expression of the fidelity using a Bures prior in
3D}\label{app-bures}

The explicit expressions of the coefficients $v^0_j$, $v^z_j$
[Eqs.~\eqref{coefficient v0j} and \eqref{coefficient vzj}] are
\begin{equation}\label{optimal V0zj}
v^0_j =2\int_{-1}^1 dr{w(r)\over
r}\left({1-r^2\over4}\right)^{J-j+{1\over2}}\left({1+r\over2}\right)^{d_j}\\
\end{equation}
and
\begin{equation}\label{Vzzj=al-be}
v^z_j=\eta_j-\nu_j ,
\end{equation}
 with
\begin{eqnarray}
&& \eta_j={d_j\over j+1}\int_{-1}^1dr{w(r)\over
r}\left({1-r^2\over4}\right)^{J-j}
 \left({1+r\over2}\right)^{d_j+1}   ,\nonumber \\
&& \nu_j= \int_{-1}^1dr{w(r)\over
r}\left({1-r^2\over4}\right)^{J-j}\!\!\!
 \left({1+r\over2}\right)^{d_j}  .
 \label{bej}
\end{eqnarray}
To obtain these expressions  we have recalled~\eqref{rhoNj} and~\eqref{rhoj-general} and
defined $w(r)=w(-r)$ for~$-1\le r<0$ to extend the
$r$-integration to the interval~$[-1,1]$.

Consider now the Bures prior~\cite{hubner:excompbures}, which is
commonly 
regarded as the natural  uniform
distribution in the Bloch sphere, since it follows
from the metric induced by the fidelity~\cite{zyk-1,petz}.
It is given by
\begin{equation}\label{drho3}
 d\rho=\frac{4}{\pi}\frac{r^2 dr}{\sqrt{1-r^2}}\,d n,
\end{equation}
which implies $w(r)=(4/\pi) r^2
(1-r^2)^{-1/2}$. In this case the integration in~\eqref{optimal
V0zj} and~\eqref{bej} can be performed analytically. For
simplicity, we will consider an even number of copies $N=2n$
($J=n$). By making extensive use of
\begin{equation}\label{beta}
    \int_{-1}^1\frac{r}{2}\left(\frac{1-r}{2}\right)^{\alpha-1}
    \left(\frac{1+r}{2}\right)^{\beta-1}=
    \frac{\beta-\alpha}{\beta+\alpha}B(\alpha,\beta),
\end{equation}
where
\begin{equation}
    B(\alpha,\beta)=\frac{\Gamma(\alpha)\Gamma(\beta)}{\Gamma(\alpha+\beta)}
\end{equation}
is the standard Euler Beta function, we obtain
\begin{equation}\label{f0-bures}
   v^0_j
    =\frac{8 d_j}{\pi(2n+3)} B(n-j+1,n+j+2).
\end{equation}
Similarly,
\begin{equation}
  \eta_j
  = \frac{8 d_j}{\pi(2n+3)}B(n-j+\smfrac12,n+j+\smfrac52) ,
\end{equation}
and
\begin{equation}
 \nu_j
  =\frac{4d_j}{\pi(2n+2)}B(n-j+\smfrac12,n+j+\smfrac32) ,
\end{equation}
which lead to
\begin{equation}
\label{f1-complete}
    v^z_j=\frac{8d_j j}{\pi}
    \frac{\Gamma(n-j+\smfrac12)\Gamma(n+j+\smfrac32)}{\Gamma(2n+4)}.
\end{equation}
Putting the various pieces together we finally obtain the closed
expression:
\begin{eqnarray}
&&\kern-3em\Delta={4\over\pi}\sum_{j=0}^n{2(2j+1)^2\over(2n+3)(2n+2)(2n+1)}\nonumber\\
&&\kern-3em\times\sqrt{ 1+\left[{j\over
n+j+1}{\Gamma(n-j+\smfrac12)\Gamma(n+j+\smfrac32)\over
\Gamma(n-j+1)\Gamma(n+j+1)}\right]^2 } . \label{Delta bures}
\end{eqnarray}

\section{Covariant POVMs for 2D~states}\label{app-covariant-povms-2d}

For the sake of completeness, in this appendix we give a simple
proof specialized to the 2D case  of a more general result
concerning the optimality of covariant (continuous)
POVMs~\cite{holevo:book}.  More precisely, we wish to prove that
for any given POVM, $\{O_\chi\}$, there is always a covariant
(continuous) one, with elements
\begin{equation}\label{covariant POVM 2D}
\tilde O_{\chi\phi}=U(\phi)\Omega_\chi U^\dagger (\phi),
\end{equation}
which gives the same average fidelity for a suitable
positive operator $\Omega_\chi$. The proof goes as follows.

In the 2D case the average fidelity can be written as (in this section the integration limits
$0$ and~$2\pi$ are understood)
\begin{equation}\label{F f(theta-thetax)}
F=\sum_\chi \int {d\theta\over2\pi}
f(\theta_\chi-\theta,R_\chi)\tr[\rho(\theta) O_\chi] ,
\end{equation}
where  $\theta_\chi$ ($\theta$) is the angle between
$\vec R_\chi$  ($\vec r$) and the $x$-axis, and
we denote the fidelity by $f(\theta_\chi-\theta,R_\chi)$ to emphasize the fact
that in~2D it is a function of  the difference of these
two angles. Note also that we drop the explicit dependence on $r$
which does not play any role in the proof. Thus,
e.g., we denote the mixed state $\rho(\vec r)$ simply as
$\rho(\theta)$.
Proving our statement amounts to proving that the POVM with
elements and associated guess given by
\begin{equation}\label{tilde O}
\tilde O_{\chi\phi}= U(\phi-\theta_\chi) O_\chi
U^\dagger(\phi-\theta_\chi) \stackrel{\mbox{\tiny\rm guess}}
{\longrightarrow}\phi, R_{\chi}
\end{equation}
gives the same fidelity as $\{O_\chi\}$. Note that~(\ref{tilde O})
defines~$\Omega_\chi$ in~(\ref{covariant POVM 2D}) through
\begin{eqnarray}
\tilde O_{\chi\phi} &=& U(\phi)\left[ U^\dagger(\theta_\chi)
O_\chi U(\theta_\chi)\right] U^\dagger(\phi)
\nonumber\\
&\equiv& U(\phi)\Omega_\chi U^\dagger (\phi).
\end{eqnarray}
In formul\ae\ we wish to prove that
$
F=\tilde F 
$,
where
\begin{equation}\label{tildeF}
\tilde F=\sum_\chi\int {d\phi\over2\pi}{d\theta\over2\pi}
f(\phi-\theta,R_\chi) \tr[\rho(\theta) \tilde O_{\chi\phi} ]
\end{equation}
is the fidelity we obtain with $\{\tilde O_{\chi\phi}\}$. We also
have to prove that $\{\tilde O_{\chi\phi}\}$ in~(\ref{tilde O}) is
indeed a POVM, namely, that
\begin{equation}
\sum_\chi\int {d\phi\over2\pi}\tilde O_{\chi\phi}=\id \label{res.
id.}
\end{equation}

Let us start by proving~(\ref{res. id.}). We simply change
variables $\phi\to\phi'=\phi-\theta_\chi$ and use the invariance
of the $\rm U(1)$ Haar measure, which in this case is the trivial
identity $\int_0^{2\pi} d\phi \,g(\phi)=\int_0^{2\pi} d\phi
\,g(\phi+\alpha)$ satisfied by any periodic function~$g$ of period~$2\pi$. 
We have
\begin{eqnarray}
\sum_\chi\int {d\phi\over2\pi}\tilde O_{\chi\phi}&=&\int
{d\phi'\over2\pi}U(\phi')\; \sum_\chi O_\chi\; U^\dagger(\phi')
\nonumber\\
&=&\int {d\phi'\over2\pi}U(\phi') U^\dagger(\phi')=\id .
\end{eqnarray}

We use the same logic to prove that $F=\tilde F$:
\begin{eqnarray}
\tilde F&=&\sum_\chi \int {d\theta\over2\pi} {d\phi\over2\pi}
f(\phi-\theta,R_\chi)
\nonumber\\
&\times& \tr[\rho(\theta)U(\phi-\theta_\chi) O_\chi
U^\dagger(\phi-\theta_\chi)]
\nonumber\\
&=& \sum_\chi \int {d\theta\over2\pi} {d\phi\over2\pi}
f(\phi+\theta_\chi-\theta,R_\chi)
\nonumber\\
&\times& \tr[\rho(\theta)U(\phi) O_\chi U^\dagger(\phi)] .
\label{tildeF=F 1}
\end{eqnarray}
We now use that
$
U^\dagger(\phi)\rho(\theta)U(\phi)=\rho(\theta-\phi) 
$
and make the change of variable~$\theta\to\theta-\phi$ to obtain~$\tilde F=F$.

If $R_\chi=R$ for all $\chi$ (this is the case if the estimation
of $r$ is entirely based on $j$, as in the last part of
Section~\ref{subsec-finite-2D}), we can replace the POVM elements
$\tilde O_{\chi\phi}$ by
\begin{equation}\label{tilde O bis}
\tilde O_\phi=\sum_\chi \tilde O_{\chi\phi}
\stackrel{\mbox{\tiny\rm guess}} {\longrightarrow}\phi .
\end{equation}
This is equivalent to
\begin{equation}\label{covariant POVM 2D bis}
\tilde O_\phi=U(\phi)\Omega U^\dagger (\phi),
\end{equation}
where the positive operator $\Omega$ can be expressed in terms of
$\Omega_\chi$ in~\eqref{covariant POVM 2D} simply as
$\Omega=\sum_\chi\Omega_\chi$. The proof that achieves the same
fidelity is straightforward and it amounts to pulling the sum over
$\chi$ into or out of the trace in Eqs.~\eqref{tildeF}
and~\eqref{tildeF=F 1}, which we are entitled to do because we are assuming
that $R_\chi$ is now independent of~$\chi$.

Using the results in Ref.~\cite{bbm-isotropic}, it is easy  to
show that for any given covariant (continuous) POVM~with elements
given by~(\ref{covariant POVM 2D}) there is always a POVM with~a
{\em finite} number of elements ${\bar O}_{\phi_a}=U(\phi_a)\Omega
U^\dagger (\phi_a)$, $a=0,1,2,\dots M-1$, which achieves the same
fidelity for a suitably large $M$. The angles $\phi_a$ can be
chosen to be $ \phi_a={2\pi a/ M}$, $a=0,1,2,\dots M-1$.


\section{Computation of the coefficients \lowercase{{\boldmath$a_j$}}
and \lowercase{{\boldmath$b_j$}}}\label{A and B} 

In this Appendix
we give an approximation to $c^j_m$, defined in~Eq.~\eqref{cjm},
of the form $c^j_m\approx a_j+b_j (m-j)$ valid for $m\approx j$ large enough.

Recalling the Wigner formula 
\begin{eqnarray}
&&\kern-3em{\rm d}^{(j)}_{mm'}(\theta)=\sqrt{(j+m)!(j-m)!(j+m')!(j-m')!}\nonumber\\
&&\kern-3em\times
\sum_{i=0}^{2j+1}{(-1)^{i}\left(\cos{\theta\over2}\right)^{2j+m'-m-2i}\left(-
\sin{\theta\over2}\right)^{m-m'+2i}\over
(j-m-i)!(j+m'-i)!(i+m-m')!i!}  , \label{wigner}
\end{eqnarray}
we obtain
\begin{eqnarray}
&&\kern-2.5em c^j_j={1\over2^{2j}}\sum_{m=-j}^j\pmatrix{2j\cr j-m}
\sqrt{{j-m\over j+m+1}},\nonumber\\
&&\kern-3.5em c^j_{j-1}=\sum_{m=-j}^j\pmatrix{2j\cr j-m}
\sqrt{{j-m\over j+m+1}}{m(1+m)\over 2^{2j-1}j} .
\label{more implicit}
\end{eqnarray}
We note that the two coefficients $c^j_j$ and $c^j_{j-1}$ are
binomial sums modulated by smooth functions of~$m$ in a neighborhood of~$m=0$. More
precisely,
\begin{equation}\label{implicit varphi}
c^j_k=\sum_{m=-j}^j\pmatrix{2j\cr j-m}{1\over2^{2j}}\varphi_k(m),
\end{equation}
where $\varphi_k(m)$, which can be read off from~\eqref{more implicit}  
for $k=j$,  $j-1$,  can be Taylor-expanded at $m=0$. For
large $j$ this expansion is
\begin{eqnarray}
\varphi_j(m)
&=&1-{1\over2j}-{m\over j}+{m^2\over2j^2}+\Or(j^{-3/2}),\nonumber\\
\varphi_{j-1}(m)
&=&{2m\over j}+\left({2\over j}-{3\over j^2}\right)
m^2\nonumber\\
&-&{2m^3\over j^2}+{m^4\over j^3} +\Or(j^{-3/2}) .
\end{eqnarray}
Here the power counting is done by noticing that $m$ is order
$\sqrt j$, since the sum
\begin{equation}
S_q=\sum_{m=-j}^j\pmatrix{2j\cr j-m}{m^q\over2^{2j}}
\end{equation}
is $\Or(j^{q/2})$ for $q$ even and vanishes  for  $q$ odd, as is
well known. In particular, we have $S_0=1$, $S_2=j/2$,
$S_4={j(3j-1)/4}$.

With all this information we obtain $c^j_{j}=1-{1/(4j)}$, $c^j_{j-1}=1-{3/(4j)}$,
and finally have
\begin{eqnarray}
c^j_{m}&=&c^j_{j}+\left(c^j_{j}-c^j_{j-1}\right)(m-j)+\Or[(m-j)^2]\nonumber\\
&=&{1\over2}\left(1-{1\over2j}\right)+{m\over2j}+\Or[(m-j)^2].
\label{cj approximate}
\end{eqnarray}


\section{Explicit computation of the asymptotic fidelity}\label{app-asymptotic-explicit}

Here we present with some detail the procedure we have used to evaluate the sum of~\eqref{n.l.o} 
in the large $N=2n$ limit. We first focus on~2D states and later
comment on the main differences with~3D. 

In the two cases, we write~$n_j$ as the right hand side of the identity
\begin{equation}
{d_j\over n+j+1}\pmatrix{2n\cr  n-j}=\pmatrix{2n\cr n-j}-
\pmatrix{2n\cr n+j+1}  .
\end{equation}

\subsection{The 2D case}

After plugging Eqs.~(\ref{v0j final}) and~(\ref{gxj after}) into Eq.~\eqref{n.l.o},
we have
\begin{eqnarray}
&&\kern-3em\Delta_{\rm2D}=\sum_{j=0}^n\left[\pmatrix{2n\cr
n-j}-\pmatrix{2n\cr
n+j+1}\right] \nonumber \\
&&\kern-2em\times
\left\{ 2\sqrt{1-{j^2\over n^2}} \int_0^1 dr {w(r)\over r}  \right.\nonumber\\
&&\kern-2em\times\left[
    \left({1-r\over2}\right)^{n-j+{1\over2}} \left({1+r\over2}\right)^{n+j+{3\over2}}
        -(r\to -r)
        \right]     \nonumber\\
&&\kern-2em+
       \int_0^1 dr {w(r)\over r}\left[
      {1\over n} \left(r j-{1\over4}\right)\right.\nonumber\\
 &&\kern-2em\times \left.  \left.   \left({1-r\over2}\right)^{n-j}
 \left({1+r\over2}\right)^{n+j+1}
       -(r\to -r)
       \right]
\right\}  . \label{messy1}
\end{eqnarray}
We next multiply the powers of $(1\pm r)/2$ that are explicitly
given in this
equation by the first binomial. Likewise, we multiply those denoted
by $(r\to -r)$ 
by the second binomial. In the resulting
expressions, we next change the summation indexes according to
$n-j=k$ and $n+j+1=k$, respectively, and  do similar changes in the
remaining crossed terms. After some algebra, we have
\begin{eqnarray}
&&\kern-2em\Delta_{\rm 2D}={1\over n}\int_0^1 dr {w(r)\over r}\left\{{1+r\over2}\right.\nonumber\\
&&\kern-1em\times\left[\sum_{k=0}^{n}\ B_k(r) \Phi_k(r)+
\sum_{k=n+1}^{2n}\ B_{k}(r) \Phi_{k-1}(r)\right]\nonumber\\
&&\kern-1em-\left.{1-r\over2}[r\to-r]\right\} , \label{sum B Phi}
\end{eqnarray}
where $B_k(r)$ is defined by
\begin{equation}\label{binom}
B_k(r)=\pmatrix{2n\cr k}\left({1-r\over2}\right)^{k}
\left({1+r\over2}\right)^{2n-k}
\end{equation}
and
\begin{equation}
\Phi_k(r)=\sqrt{k(2n-k)(1-r^2)}+(n-k)r-{1\over4} .
\end{equation}
Since the coefficients $B_k(r)$ are the terms of a binomial
series, for large $n$ only those for which $k\approx n(1-r)\le n$
(or equivalently $2n-k\ge n$) give a significant contribution to
the fidelity, whereas the rest fall off exponentially with~$n$.
This enables us to expand the factor $\sqrt{(k-1)(2n-k+1)}$ in $\Phi_{k-1}(r)$ as a power series in
$1/k$ and~$1/(2n-k)$ and obtain the relation
\begin{eqnarray}
&&\kern-2em\Phi_{k-1}(r)=\Phi_k(r)+{\sqrt{1-r^2}\over2}\nonumber\\
&&\kern-1em\times\left(\sqrt{{k\over2n-k}}-\sqrt{{2n-k\over
k}}\;\right)+r+\ord(1/n) \label{Phik-1}
\end{eqnarray}
which we use in the second sum of~\eqref{sum B Phi}. We further define
$\Psi_k(r)=\Phi_{k-1}(r)-\Phi_k(r)+\ord(1/n)$. It satisfies
$\Psi_k(r)=-\Psi_{2n-k}(-r)$, as can be read off from~\eqref{Phik-1}.

The leading contributions come from the terms 
that contain $\Phi_k(r)$, and the corresponding term in
$[r\to -r]$. They combine into a single sum from $k=0$ to $k=2n$. The
rest of the terms [those proportional to $\Psi_{k}(r)$ and
$\Psi_{k}(-r)$] are subleading 
and can be
simplified using the change of indexes $k\to 2n-k$. The result can
be cast as
\begin{eqnarray}
&&\kern-2em\Delta_{\rm 2D}={1\over n}\int_0^1 dr {w(r)\over r}\left\{
\left[{1+r\over2}\sum_{k=0}^{2n}\ B_k(r) \Phi_k(r)
\right.\right.\nonumber\\
&&\left.\left.\kern-1em+ {1-r\over2}\sum_{k=0}^{n-1}\ B_{k}(r)
\Psi_{k}(r)\right] -[r\to-r]\right\} . \label{sum B Phi bis}
\end{eqnarray}
We readily see that the first sum (as well as the corresponding
one obtained by the substitution $r\to -r$) is a binomial sum
modulated by the function $\Phi_k(r)$, analogous
to~\eqref{implicit varphi} in Appendix~\ref{A and B}, and can be
computed along the same line. This sum is peaked at $k\approx
n(1-r)$, as we have already mentioned, which suggests expanding
$\Phi_k(r)$ in powers of $k-n(1-r)$. More precisely, one can check
that
\begin{equation}
\Phi_k(r)=n-{1\over4}-{[k-n(1-r)]^2\over2n(1-r^2)}+\ord(1/n)
\end{equation}
[the power counting is simply $k-n(1-r)=\Or(\sqrt n)$]. Recalling that
the lowest moments, $S_q(r)=\sum_{k=0}^{2n} B_k(r)[k-n(1-r)]^q$, of the binomial series given
by~(\ref{binom}) are
$S_0(r)=1$, $S_2(r)=(n/2)(1-r^2)$ we obtain
\begin{equation}
    \sum_{k=0}^{2n}B_k(r)\Phi_k(r)=n-1/2+\ord(1/n).
\end{equation}

To evaluate the second sum in~\eqref{sum B Phi bis} we use again
the approximation $B_k(r)\to \delta[k-n(1-r)]$ [see Eq.~\eqref{gaussian} and
the comments below it], along with the substitution
$\sum_{k=0}^{n-1}\to n\int_0^1 ds$, where $s=k/n$. This yields
\begin{eqnarray}
&&\kern-2em\sum_{k=0}^{n-1}B_k(r)\Psi_k(r)=
\Or(1/n)  .
\label{BPsi}
\end{eqnarray}
The counterpart of~\eqref{BPsi}  in the term denoted by~$[r\to-r]$,
Eq.~\eqref{sum B Phi bis}, gives no contribution since
$\delta[k-n(1+r)]$ lies outside the $s$-integration range.
Collecting the various pieces we finally obtain
\begin{equation}
\Delta_{\rm2D}=\left(1-{1\over2n}\right)\int_0^1 dr \,w(r)+\ord(1/n).
\end{equation}

\subsection{The 3D case}

The 3D case is quite similar. Our starting point is now Eqs.~\eqref{optimal V0zj} 
and~\eqref{Vzzj=al-be}. We proceed as above to obtain 
\begin{eqnarray}
    \nonumber
    v_j^z&=&\int_{-1}^1 dr\frac{w(r)}{r}\left\{\left[\frac{(2j+1)r-1}{2(j+1)}\right]\left(\frac{1-r}{2}
    \right)^{n-j}\right.\\
    &\times&\left.\left(\frac{1+r}{2}\right)^{n+j+1}-(r\rightarrow-r)\right\}
    \label{seeforinstance1}
\end{eqnarray}
and a similar expression for~$v_j^0$. From them we compute~$\Delta_{\rm 3D}$ to be
\begin{eqnarray}
&&\kern-2.7em\Delta_{\rm 3D}\!=\!
    \frac{1}{n}\sum_{j=0}^n\!\left[\pmatrix{2n\cr n-j}\!-\!\pmatrix{2n\cr n+j+1}\right] \!
     \int_{-1}^1\!\! \! \! dr\frac{w(r)}{r}\nonumber\\
    &&\kern-2em\times\Bigg\{\left[\sqrt{(n^2-j^2)(1-r^2)}+\frac{(2j+1)j r-j}{2(j+1)}\right]
 \nonumber\\
    &&\kern-2em\times \left(\frac{1-r}{2}\right)^{n-j}
    \left(\frac{1+r}{2}\right)^{n+j+1}-[r\rightarrow-r]\Bigg\}.
    \label{seeforinstance2}
\end{eqnarray}
This expression can be cast in the form of~\eqref{sum B Phi}, where now
\begin{eqnarray}
  \Phi_k(r)&=&\sqrt{k(2n-k)(1-r^2)} +(n-k)r \nonumber \\[.5em]
   &-&{1\over2}\frac{\sqrt{n^2-k(2n-k)}}{\sqrt{n^2-k(2n-k)}+1} .
\end{eqnarray}
One can check that $\Psi_k(r)$ is again defined by~\eqref{Phik-1} 
and $\Delta_{\rm 3D}$ can thus be expressed in the form~\eqref{sum B Phi bis}. 
The first sum is
again Taylor-expanded about $k=n(1-r)$. Using the moments
of the binomial series defined by~(\ref{binom}) 
and keeping only the relevant order we obtain
\begin{equation}
    \sum_{k=0}^{2n}B_k(r)\Phi_k(r)=n-\frac{3+2|r|}{4}+\Or(1/n)   .
\end{equation}
Note that we cannot drop the absolute value since the integral over $r$ extends
to the interval~$[-1,1]$ [see, e.g. Eqs.~(\ref{seeforinstance1}) and~(\ref{seeforinstance2})].

To evaluate the second sum in~\eqref{sum B Phi bis} we proceed as in the previous 2D~case, and find that~\eqref{BPsi} still holds.
Finally, we obtain 
\begin{equation}
    \Delta_{\rm 3D}=\int_0^1 dr\,w(r)\left(1-\frac{3+2r}{4n}\right)+\ord(1/n)  .
\end{equation}


\section{SLDs  and {\boldmath $C^H[H(\tht_0)]$} for the 3D model\label{sec:applambdas}}
The SLDs of the 3D model can be calculated to be
\beq
\lambda_r\!\!&=&\!\!\frac{1}{1+r}\frac{\id+\vec{n}\cdot\vec{\sigma}}{2}-\frac{1}{1-r}
\frac{\id-\vec{n}\cdot\vec{\sigma}}{2},\\[.5em]
\lambda_\theta\!\!&=&\!\!r\,\partial_\theta \vec{n} \cdot \vec{\sigma}, \\[.5em]
\lambda_\phi\!\!&=&\!\! r\,\partial_\phi \vec{n} \cdot
\vec{\sigma}.
\eeq
[In this appendix we drop the arguments $\tht=(r,\theta,\phi)$ and~$\tht_0$ wherever no confusion arises.]
The two SLD of the 2D model, $\lambda_r$ and $\lambda_\theta$,
are obtained by simply setting~$\theta=\pi/2$ and then replacing~$\phi$ by~$\theta$ in the above expressions.

To compute  $C^H(H)$ we first need $\xt=(X_r,X_\theta,X_\phi)$,
 which are completely fixed by the conditions
\begin{eqnarray}
  X_\alpha &=& X_\alpha^{\dagger}, \label{F1}\\[.5em]
  \tr \rho X_\alpha &=& 0 \label{F2}\\[.5em]
  \tr \partial_\alpha \rho X_\beta  &=& \delta_{\alpha \beta}
  \label{F3}
\end{eqnarray}
Hermiticity, Eq.~(\ref{F1}), requires
\begin{equation}
X_\alpha= a_\alpha \id +  \vec{b}_\alpha \cdot
\vec{\sigma},\quad \alpha=r,\theta,\phi   .
\end{equation}
The conditions \eqref{F2} yield
 \beq \label{eq:firstcond}
 a_\alpha+ \vec{b}_\alpha\cdot \vec{n}=0,
 \eeq
and conditions \eqref{F3} give
\beq
 \vec{b}_r&=&\vec{n}, \\[.5em]
 \vec{b}_\theta&=&\frac{1}{r}\,\partial_\theta \vec{n},\\[.5em]
 \vec{b}_\phi&=&\frac{1}{r\sin^2\theta}\,\partial_\phi \vec{n}.
\eeq
These together with~(\ref{eq:firstcond}) imply $a_r=-r$,
$a_\theta=0$, and~$a_\phi=0$. Hence,  the only set of matrices
satisfying~(\ref{F1}-\ref{F3})~is
 \beq
 X_r&=&-r\,\id+\vec{n}\cdot\vec{\sigma}, \nonumber\\
 X_\theta&=&\frac{1}{r}\, \partial_\theta \vec{n}\cdot\vec{\sigma},\\
 X_\phi&=&\frac{1}{r\sin^2\theta}\,\partial_\phi \vec{n}
 \cdot\vec{\sigma}.  \nonumber
 \eeq

To compute the Holevo bound we only need to take traces of the
form~$\tr\rho X_\alpha X_\beta$. A straightforward calculation
gives
 \beq
\rlprt Z[\xt]&=&H_{\rm 3D}^{-1},   \\[1em]
\imprt Z[\xt]&=& \left( \begin{array}{@{}ccc@{}}
0&0&0\\
0&0&\displaystyle \frac{1}{r \sin\theta}\\
0&\displaystyle -\frac{1}{r \sin\theta}&0
\end{array}\right).
\eeq
Therefore
\beq
\tr H_{\rm 3D}H_{\rm 3D}^{-1}&=&3,\\[.5em]
\tr \left|\sqrt{H_{\rm 3D}}\imprt Z[\xt]\sqrt{H_{\rm 3D}}\,\right|&=&2r, 
\eeq 
and we obtain~(\ref{C^H 3D}).

\section{Van Trees asymptotic bound for 2D
states}\label{app:vantrees}

 \newcommand{\thetabf}{\boldsymbol\theta}
\newcommand{\psibf}{\boldsymbol\psi}

Let $\thetabf$ be the column vector of the two real parameters~$r$ 
and~$\theta$ of Sec.~\ref{subsec-finite-2D},
which we use to parametrize  the states on the equatorial plane of
the Bloch sphere. Define $\psibf( \thetabf)=\frac 12
\rbf(\thetabf)$ where $\rbf$ is the four-dimensional real vector
(of length $1$) introduced in Sec.~\ref{sect-preli}. By~(\ref{f-qubits}) we can now
write
\begin{equation}
1-f(\thetabf_0,\htht) = \| \psibf(
\thetabf_0)-\psibf(\htht)\|^2
\end{equation}
showing that one minus the fidelity is the squared $L_2$ cost
function for estimating $\psibf$. Taking the two states close to
one another, and comparing with~(\ref{H from f}) shows that
\begin{equation}
\psibf'^\top\psibf'=\frac14 H
\end{equation}
where $\psibf'(\thetabf)$ denotes the $4\times 2$ matrix of
partial derivatives of $\psibf$ with respect to components of
$\thetabf$ and $H$ is the QFI.

Let $\bar I^N= I^N/N$ denote the normalized FI 
for $\thetabf$ based on an arbitrary collective measurement
on the $N$ copies, and let $\htht$ denote an arbitrary
estimator of $\thetabf$ based on that measurement. By $\mathbb
E_w$ we denote averaging over $\thetabf$ with respect to a prior
probability density~$w$ over the equatorial plane. Then the
 van Trees inequality~\cite{vantrees}  states that,
for any given matrix function $C(\thetabf)$ of size
 $\mathrm {dim}(\psibf)\times\mathrm{dim}(\thetabf)$,
and under certain smoothness conditions on the probability
distribution of the outcome of the measurements and on the prior~$w$,
\begin{eqnarray}
&& N\mathbb E_w  \| \psibf( \thetabf_0)-\psibf(\htht)\|^2 ~\ge~
\nonumber\\[.5em]
&&~~~~ \frac{(\mathbb E_w\, \tr C\psibf'^\top)^2}
 {\mathbb E_w\, \tr C \bar I^N C^\top +
{\displaystyle1\over \displaystyle N}
 \mathbb E_w{\displaystyle \frac {(wC)'^\top  (wC)'} {w^2}} } ,
\end{eqnarray}
 where by $(wC )'$ we denote the column vector
of the same length as  $\psibf$,
 with row elements $\sum_\beta \partial_\beta [w(\thetabf) C_{ i \,\beta}(\thetabf)]$.
By~the Helstrom information inequality~(\ref{eq:IleqH}) we may bound $\bar
I^N$ in the denominator by $H$ (of the single-copy model). Without the ``$1/N$'' term in the
denominator, the optimal choice of $C$ would be $C=\psibf' H^{-1}$.
Making this choice anyway gives
 \begin{eqnarray}\label{vtr1}
\kern-2em&&  N\mathbb E_w  \| \psibf( \thetabf_0)-\psibf(\htht)\|^2 ~\ge~
\nonumber \\[.5em]
\kern-2em&&
 \frac{(\mathbb E_w\, \tr \psibf' H^{-1}\psibf'^\top)^2}
 {\mathbb E_w\, \tr  \psibf' H^{-1} H H^{-1}{\psibf'}^\top  +{
\displaystyle1\over \displaystyle N}
 \mathbb E_w {\displaystyle \frac {(wC)'^\top  (wC)'} {w^2} }}  .
 \end{eqnarray}
 Hence,
 provided the second term in the denominator is finite,
 by further substituting $\psibf'^\top\psibf'=\frac14 H$ and letting~$N$
 converge to infinity, we obtain
 \begin{equation}\label{vtr2}
 \mathop{\lim\inf}\limits_{N\to\infty} N\mathbb E_w \mathbb E_{\thetabf}  (1-f(\thetabf,\htht)) ~\ge~
 \frac12.
 \end{equation}

 The van Trees inequality requires some modest smoothness of the probability density of
 the measurement outcomes as function of $\thetabf$, which are satisfied in our case
 since the density matrix
 $\rhoN(\thetabf)$ is a smooth function of~$\thetabf$. It requires smoothness of the
 prior density $w$ and also that this density converges to zero at the boundary of its
 support. This last property does not hold for the priors in which we are interested.
 However, for a given prior $w$ and for given $\epsilon>0$ one can construct
 a prior $w_\epsilon$ which is zero outside a circle of radius
 strictly smaller than $1$,
 which converges
 smoothly to zero at the boundary of its support, and which is everywhere smaller than
 $(1+\epsilon)w$. The  modification of $w$ can simultaneously be done  ensuring
 that the second term in the denominator of (\ref{vtr1}) is finite. Since
  \begin{equation}
\mathbb E_w \mathbb E_{\thetabf}
(1-f(\thetabf,\htht)) ~\ge~
 \frac{ \mathbb E_{w_\epsilon} \mathbb E_{\thetabf}  (1-f(\thetabf,\htht))}{1+\epsilon}
  \end{equation}
 we can first derive (\ref{vtr2}) with $w$ replaced by $w_\epsilon$, then let~$\epsilon\to 0$,
 resulting in (\ref{vtr2}) with the original $w$ in place.


\end{document}